\documentclass[aps,prd,showpacs,nofootinbib]{revtex4}
\usepackage{amsmath,amssymb,graphicx}
\usepackage{epsf}
\usepackage{bm}

\newcommand{\nc}{\newcommand}
\nc{\postscript}[2]
{\setlength{\epsfxsize}{#2\hsize}\centerline{\epsfbox{#1}}}
\nc{\non}{\nonumber}
\nc{\hc}{\hbox {h.c.}} \nc{\re}{\hbox {Re}} 
\nc{\mev}{\hbox {MeV}} \nc{\gev}{\;\hbox {GeV}} \nc{\tev}{\;\hbox {TeV}}
\def\lsim{\mathrel{\raise.3ex\hbox{$<$\kern-.75em\lower1ex\hbox{$\sim$}}}}
\def\gsim{\mathrel{\raise.3ex\hbox{$>$\kern-.75em\lower1ex\hbox{$\sim$}}}}

\nc{\etal}{{\it et al.}}
\nc{\Lsp}{\;\;\;\;\;\;\;\;\;\;}  \nc{\LLLsp}{\lspace \lspace}
\nc{\lsp}{\;\;\;\;\;\;}
\nc{\spac}{\;\;\;}
\nc{\noi}{\noindent}
\nc{\beq}{\begin{equation}}   \nc{\eeq}{\end{equation}}
\nc{\bea}{\begin{eqnarray}}   \nc{\eea}{\end{eqnarray}}
\nc{\baa}{\begin{array}}      \nc{\eaa}{\end{array}}
\nc{\bit}{\begin{itemize}}    \nc{\eit}{\end{itemize}}
\nc{\ben}{\begin{enumerate}}  \nc{\een}{\end{enumerate}}
\nc{\bce}{\begin{center}}     \nc{\ece}{\end{center}}

\def\calQ{{\cal Q}}
\def\calD{{\cal D}}

\def\sq2{\sqrt{2}}

\def\ph{\varphi}

\def\m4{m^4(\ph)}
\def\mn2{m_n^2}

\def\v5{V^{(5)}}

\begin{document}

\title{\begin{flushright}
  \mbox{\normalsize \rm UMD-PP-09-039}
  \end{flushright}
  \vskip 20pt
Higgs Mediated FCNC's in Warped Extra Dimensions}
\author{Aleksandr Azatov}
\author{Manuel Toharia}
\author{Lijun Zhu}
\affiliation{Maryland Center for Fundamental Physics, Department of
  Physics, University of Maryland\\ College Park, MD 20742, USA}

\date{\today}

\begin{abstract}

In the context of a warped extra-dimension with Standard Model
fields in the bulk, we obtain the general flavor structure of the
Higgs couplings to fermions. These couplings will be generically
misaligned with respect to the fermion mass matrix, producing large
and potentially dangerous flavor changing neutral currents (FCNC's).
As recently pointed out in [arXiv:0906.1542], a similar effect is
expected from the point of view of a composite Higgs sector, which
corresponds to a 4D theory dual to the 5D setup by the AdS-CFT
correspondence. We also point out that the effect is independent of
the geographical nature of the Higgs (bulk or brane localized), and
specifically that it does not go away as the Higgs is pushed towards
the IR boundary. The FCNC's mediated by a light enough Higgs
(specially their contribution to $\epsilon_K$) could become of
comparable size as the ones coming from the exchange of Kaluza-Klein
(KK) gluons. Moreover, both sources of flavor violation are
complementary since they have inverse dependence on the 5D Yukawa
couplings, such that we cannot decouple the flavor violation effects by
increasing or decreasing these couplings.
We also find that for KK scales of a few TeV, the Higgs couplings to
third generation fermions could experience suppressions of up to
$40\%$ while the rest of diagonal couplings would suffer much milder
corrections. Potential LHC signatures like the Higgs flavor violating
decays $h\to\mu\tau$ or $h\to tc$, or the exotic top decay channel
$t\to c h$, are finally addressed.

\end{abstract}

\maketitle
%%%%%%%%%%%%%%%%%%%%%%%%%%%%%%%%%%%%%%%%%%%%%%%%%%%%%%%%%%%%%%%%%%%%%%%%%%%%%%%%%%%%%%%%%%%%%%%%%%%%%
\section{Introduction}

Introducing a warped extra-dimension in such a way as to create an
exponential scale hierarchy between the two boundaries of the extra
dimension~\cite{RS1} has generated a lot of attention in the recent
years as a novel approach to solve the hierarchy problem. By placing
the Standard Model (SM) fermions in the bulk of the extra dimension
it was then realized that one can simultaneously address the flavor
hierarchy puzzle of the SM~\cite{a,bulkSM}. The electroweak
precision tests put important bounds on the scale of new physics
but by introducing custodial symmetries \cite{RSeff} one can have it
around few TeV \cite{RSeff,AgashePGB,EWPTmodel,Agashezbb}.

In this paper we will study the class of models in which all the SM
fields are in the bulk and the hierarchies in masses and mixings in
the fermion sector are explained by small overlap integrals between
fermion wave functions and the Higgs wave function along the extra
dimension. This scenario can lead to the observed fermionic masses
without any hierarchies in the initial 5D Lagrangian, so that our
fundamental 5D Yukawa couplings have no structure and are all of
the same order. Another interesting feature of these models is that
the contributions to low energy observables coming from the exchange
of heavy KK states will be suppressed by the so called ``RS GIM''
mechanism \cite{AgashePerezSoni,hs}. In spite of it, it was still
found that $\Delta F=2$ processes push the mass of the KK
excitations to be above $\sim 10$ TeV
\cite{Weiler,BlankedeltaF2,Fitzpatrick:2007sa,Davidson:2007si},
making it very hard to produce and observe them at the LHC
\cite{kkgluon,Agashe:2007ki}. These bounds coming from flavor
violation in low energy observables can be avoided by introducing
additional flavor symmetries
\cite{Fitzpatrick:2007sa,Davidson:2007si,Cacciapaglia:2007fw,Csakifp}.
Another way to relax these low energy constraints is to promote the
Higgs to be a 5D bulk field (instead of being brane localized). In
this situation the bounds from $\epsilon_k$ could allow masses of
the lowest KK gluon to be as low as $\sim3$ TeV, although combining
this result with the bounds from dipole moment operators
$(b\rightarrow s \gamma, s\to d g)$ pushes back the KK scale to be
above $\sim 5$ TeV \cite{Agashe2site, Gedalia:2009ws}. A similar
tension was found in the lepton sector in \cite{Agashe:2006iy}.

It has recently been pointed out that in the context of a composite
Higgs sector of strong dynamics, one generically
expects some amount of flavor changing neutral currents (FCNC's)
mediated by the Higgs \cite{Agashe:2009di} (from an effective field
theory point of view see also the earlier works
\cite{Buchmuller:1985jz,delAguila:2000aa,Babu:1999me,Giudice:2008uua}).
In the 5D picture, the presence of KK fermion states will actually produce a misalignment
between the Higgs Yukawa couplings and the SM fermion masses, giving rise to
tree-level flavor violating couplings of the Higgs to fermions. The
induced FCNC's are strongly constrained by various low energy
experiments; if these constraints are somehow evaded, interesting
signals at the LHC could also be generated.

The possibility of a flavor misalignment between the Higgs Yukawa
matrices and the fermion mass matrices in the context of 5D warped
scenarios was first briefly mentioned in \cite{Agashe:2006wa}, although
it was not until \cite{NeubertRS} where a detailed analysis of the
flavor structure of the couplings of the Higgs (brane localized) was
first performed. There, the effects on the flavor violating Higgs
couplings were found to be small (except for third generation
quarks), with the (hidden) assumption that the contribution from a
specific type of operators is negligible. In a more general Higgs
context (bulk or brane localized), all the sources of Higgs flavor
violation were then pointed out in \cite{BlankedeltaF2,Gori},
including the previously neglected operators, although no analysis
on the overall size of the Higgs FCNC's was performed. Moreover, in
the limit of a brane localized Higgs, the effects of the larger
sources of flavor are claimed to become negligible, and so it is
again found that Higgs mediated FCNC's are highly suppressed in the
case of a brane Higgs.

In this work, we show that the induced misalignment in the Higgs
couplings is generically large and phenomenologically important in
both bulk and brane localized Higgs scenarios. The main cause for
this result is the effect of the originally neglected operators
which, due to a subtlety in the treatment of the brane localized
Higgs, ends up surviving in the brane limit and giving rise to
important misalignments between the Higgs Yukawa couplings and the
fermion mass matrices.

The outline of the paper is as follows: in section II we review the
model independent argument such that (TeV suppressed) higher
order effective operators in the Higgs sector can lead to potentially
large Higgs FCNC's.
This is then applied to the 5D RS model, first in the mass insertion approximation
in order to quickly estimate the size
of the corrections. In section III we proceed with
a more precise calculation of the Higgs Yukawa couplings in the
case of one fermion generation, and for a bulk Higgs scenario.
The deviation in the Yukawa couplings is quite
insensitive to how much the Higgs is localized near the IR brane; this
result is confirmed in section IV by doing a 5D computation for
the case of an exactly IR localized Higgs field, and it seems at odds
with the mass insertion approximation which suggests that the
corrections to the flavor violating Higgs couplings should vanish in the brane
Higgs limit. This apparent contradiction is addressed and resolved in
that same section. In section V we extend our results to the case of three
generations and then in section VI, we give an estimate of the
expected overall size of the Yukawa coupling matrices. We also argue that the couplings of the
Higgs to third generation fermions might be significantly
suppressed. These estimates are confirmed in section VII by the
results of our numerical scan. Finally, section VIII is devoted to
the study of phenomenological implications of Higgs mediated flavor
violations, where we discuss low energy bounds arising from $\Delta
F=2$ processes as well as interesting collider signatures.

%%%%%%%%%%%%%%%%%%%%%%%%%%%%%%%%%%%%%%%%%%%%%%%%%%%%%%%%%%%%%%%%%%%%%%%%%%%%%%%%%%%%%%%%%%%%%%%%%%%%%
\section{Flavor misalignment estimate}\label{diagrammatic}

From an effective field theory approach it is easy to write the lowest
order operators responsible for generating a misalignment in flavor
space between the Higgs Yukawa couplings and the SM fermion masses.
For simplicity we focus on the down quark sector and write the
following dimension 6 operators of the 4D effective Lagrangian
\cite{Buchmuller:1985jz,delAguila:2000aa,Babu:1999me,Giudice:2008uua,Agashe:2009di}:
\bea
\lambda_{ij} \frac{H^2}{\Lambda^2}\ H\overline{Q}_{L_i}\!
D_{R_j},\hspace{.5cm}  k^D_{ij}
\frac{H^2}{\Lambda^2}\ \overline{D}_{R_i}\partial\hspace{-.18cm}/
D_{R_j}\hspace{.5cm}{\rm and}\hspace{.5cm} k^Q_{ij}
\frac{H^2}{\Lambda^2}\ \overline{Q}_{L_i} \partial\hspace{-.18cm}/
Q_{L_j}
\eea
where $Q_{L_i}$ and $D_{R_j}$ are the fermionic $SU(2)$ doublets and
singlets of the SM, with $\lambda_{ij}$, $k^D_{ij}$ and $k^Q_{ij}$
being complex coefficients and $i,j$ are flavor indices; $\Lambda$ is
the cut-off or the threshold scale of the effective Lagrangian.
Upon electroweak symmetry breaking (EWSB), these operators will give a
correction to the fermion kinetic terms and to the fermion mass terms.
Calling $y_{ij}$ the original Yukawa couplings, the corrected fermion
mass and kinetic terms become:
\bea
v_4\left(y_{ij}+\lambda_{ij} \frac{v_4^2}{\Lambda^2}\right)
\overline{Q}_{L_i}\! D_{R_j},
\hspace{.5cm}  \left(\delta_{ij}/2+ k^D_{ij}
\frac{v_4^2}{\Lambda^2}\right)
\overline{D}_{R_i}\partial\hspace{-.18cm}/ D_{R_j}
\hspace{.5cm}
{\rm and}\hspace{.5cm}
\left(\delta_{ij}/2+k^Q_{ij}\frac{v_4^2}{\Lambda^2}\right)
\overline{Q}_{L_i}\partial\hspace{-.18cm}/ Q_{L_j},
\label{masskinetic}
\eea
where $v_4=174$ GeV is the Higgs electroweak $vev$, i.e. $H=h/\sqrt{2}+v_4$,
with $h$ being the physical Higgs scalar. On the other hand, the induced
operators involving two fermions and one physical Higgs $h$ become:
\bea
\left(y_{ij}+3\lambda_{ij} \frac{v_4^2}{\Lambda^2}\right)
\frac{h}{\sqrt{2}}\overline{Q}_{L_i}\! D_{R_j},
\hspace{.7cm}  \left(2 k^D_{ij}
\frac{v}{\Lambda^2}\right)
\frac{h}{\sqrt{2}}\overline{D}_{R_i}\partial\hspace{-.18cm}/ D_{R_j}
\hspace{.7cm}{\rm and}\hspace{.5cm}
\left(2k^Q_{ij}\frac{v_4}{\Lambda^2}\right)
\frac{h}{\sqrt{2}}\overline{Q}_{L_i}\partial\hspace{-.18cm}/ Q_{L_j}.
\label{yukawakinetic}
\eea
From Eq.(\ref{masskinetic}) it is clear that one has to redefine the
fermion fields to canonically normalize the new kinetic terms and then
perform a bi-unitary transformation to diagonalize the resulting mass
matrix. These fermion redefinitions and rotations will not in general
diagonalize the couplings from Eq.~(\ref{yukawakinetic}) and
therefore, we will obtain tree-level flavor changing Higgs couplings, with
a generic size controlled by $\frac{v^2}{\Lambda^2}$.

In the warped extra dimensions scenarios that we are interested in, we
can estimate easily the size of this type of misalignments between the
Higgs Yukawa couplings and the SM fermion masses by using the insertion
approximation in KK language.
The 5D spacetime we consider takes the usual Randall-Sundrum
form~\cite{RS1}:
\bea ds^2 = \frac{1}{(kz)^2}\! \Big(\eta_{\mu\nu}
dx^\mu dx^\nu -dz^2\Big), \label{RS}
\eea
with the UV (IR) branes localized at $z = R$ ($z = R^\prime$) and with
$k$ being the curvature scale of the AdS space. We are interested here in the
flavor structure of the Yukawa couplings between the Higgs and the
fermions. However, it is instructive to first consider the case of
only one generation and study the (potentially large) corrections
induced to the single Yukawa coupling. One can then easily generalize to
three generations and find the misalignment between the fermion mass
matrix and the Yukawa couplings matrix.

\begin{figure}
\begin{center}
\includegraphics[scale = 0.9]{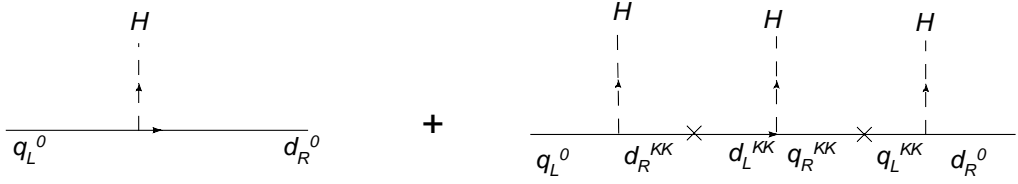}
\caption{Shift in masses and Yukawa couplings of SM fermions using
the mass insertion approximation.}\label{insertion}
\end{center}
\end{figure}

We will focus on the down-quark sector of a simple setup in which
we consider the 5D fermions $Q$, $D$. They contain the 4D SM
$SU(2)_L$ doublet and singlet fermions respectively with a 5D action
\bea
\label{fermionaction} &&\hspace{-.5cm}
S_{\text{fermion}}\!=\!\int d^4x dz \sqrt{g} \Big[ {i \over 2}
\left(\bar{Q} \Gamma^A {\cal D}_A Q -
{\cal D}_A \bar{ Q} \Gamma^A Q\right)  %\non&&\hspace{-.5cm}
+ {c_{q} \over R} \bar{ Q} {Q} + (Q\rightarrow D)
+\left(Y_d\ \bar{Q} { H} D + h.c.\right) \Big]
\eea
where $c_{q}$ and $c_{d}$ are the 5D fermion mass coefficients and
$ H$ is the bulk Higgs field localized towards IR brane. The
wavefunctions of the fermion zero modes are determined by their
corresponding 5D mass coefficients. To obtain a chiral spectrum, we
choose the following boundary conditions for $Q, D$
\begin{eqnarray}
Q_L (+ +),\quad Q_R(- -), \quad D_L(- -), \quad D_R(+ +).
\end{eqnarray}
Then, only $Q_L$ and $D_R$ will have zero modes, with wavefunctions:
\begin{eqnarray}
q_L^0(z) &=& f(c_q)\frac{{R'}^{-\frac{1}{2}+c_q}}{ R^{2}} z^{2-c_q}\\
d_R^0(z) &=& f(-c_d)\frac{{R'}^{-\frac{1}{2}-c_d}}{ R^{2}} z^{2+c_d},
\end{eqnarray}
where we have defined $f(c) \equiv
\sqrt{\frac{1-2c}{1-\epsilon^{1-2c}}}$ and the hierarchically small
parameter $\epsilon=R/ R'\approx 10^{-15}$, which is generally
referred to as the ``warp factor''. Thus, if we choose $c_q (-c_d) > 1/2$,
then the zero modes wavefunctions are localized towards the UV brane; if
$c_q (-c_d) < 1/2$, they are localized towards the IR brane. The
wavefunctions of the KK modes are all localized near the IR
brane. Note that the wavefunctions of the KK modes $Q_R$ and $D_L$
vanish at the IR brane due to their boundary conditions. The Yukawa
couplings of the Higgs with fermions (zero modes or heavy KK modes)
are set by the overlap integrals of the corresponding
wavefunctions. For a bulk Higgs localized near the IR brane,
the zero-zero-Higgs, zero-KK-Higgs, KK-KK-Higgs Yukawa couplings are
given approximately by
\begin{eqnarray}
Y_{d,00} &\sim& Y_* f(c_q) f(-c_d)\\
Y_{d, 0n} &\sim& Y_* f(c_q) \, \, \text{or}\,\, Y_* f(-c_d)\\
Y_{d, nm} &\sim& Y_*
\end{eqnarray}
where $Y_* = Y_d /\sqrt{R}$ is the $O(1)$ dimensionless 5D Yukawa
coupling, and we ignored $O(1)$ factors in the equations above. The
SM fermions are mostly zero mode fermions with some small amount of
mixing with KK mode fermions. Therefore, we can use the mass
insertion approximation to calculate the masses and Yukawa couplings
of SM fermions. This is shown in Fig. \ref{insertion}, where
$q^0_L$, $d^0_R$ are zero modes of $SU(2)_L$ doublet and singlet
fermions respectively and $q^{KK}_L$, $q^{KK}_R$, $d^{KK}_L$,
$d^{KK}_R$ are KK mode fermions. From the Feynman diagram in Fig.
\ref{insertion} we see that the SM fermion mass is given by
\begin{eqnarray}
m^d_{SM} &\approx& Y_{d,00}\ v_4 - Y_{d,0n} Y_{d,nm} Y_{d,m0}\ v_4
\frac{v^2}{M_{KK}^2}\\ \nonumber &\approx&
f(c_q) Y_* f(-c_d)\ v_4 - f(c_q) \frac{Y_*^2 v_4^2}{M_{KK}^2} f(-c_d)
Y_*\ v_4
\end{eqnarray}
where $v_4$ is the Higgs vev and we assume that all KK fermion
masses are of the same order ($M_{KK}$).

The 4D effective Yukawa couplings of SM fermions can be calculated
using the same diagram. However in the second diagram of
Fig. \ref{insertion}, we have to set two external $H$ to their vev
$v_4$ while the other one becomes the physical Higgs $h$, and there
are three different ways to do this. Thus we obtain the 4D Yukawa couplings
\begin{equation}
y^d_{SM} \approx f(c_q) Y_* f(-c_d)  - 3 f(c_q) \frac{Y_*^2v_4^2}{M_{KK}^2} f(-c_d) Y_*
\end{equation}
We see that the SM fermion masses and the 4D Yukawa couplings are not
universally proportional; indeed there is a shift with respect to the
SM prediction of  $m^d_{SM}=y^d_{SM}v_4$.

We thus define the shift $\Delta^d$ as
\bea
\Delta^d &=&  m^d_{SM} -\ y^d_{SM}v_4
\eea
and it is easy to see that the contribution of the diagrams of
Fig. \ref{insertion} to $\Delta^d$ is
\bea
\label{delta1}
\Delta_1^d &\approx& 2f(c_q) \frac{Y_*^2 v_4^2}{M_{KK}^2} f(-c_d)v_4 Y_*.
\eea

\begin{figure}
\begin{center}
\includegraphics[scale = 0.9]{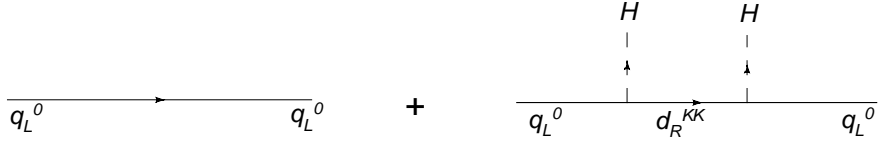}
\caption{Correction to kinetic terms using insertion
approximation.}\label{insertion1}
\end{center}
\end{figure}

There is yet another source of shift between masses and Yukawa couplings
coming this time from the corrections to the kinetic terms. This
is the contribution which was pointed out and carefully computed in
\cite{NeubertRS}, and as wee will see later, in agreement with our own
results for that specific term. As shown in Fig. \ref{insertion1}, the
kinetic term for the fermion mode $q^{SM}_L$ receives a correction
induced by the mixing with KK fermion modes
\begin{equation}
\left(1 +  Y_{d,0n} Y_{d,n0}
\frac{H^2}{M_{KK}^2}\right)\bar{q}^{SM}_L i {\partial}\hspace{-.18cm} / q^{SM}_L
\approx\left(1 +  f(c_q)^2 \frac{(Y_*
H)^2}{M_{KK}^2}\right)\bar{q}^{SM}_L i{\partial}\hspace{-.18cm} / q^{SM}_L
\end{equation}
After redefining fields so that their kinetic term is
canonical, there will be a new contribution to the shift between
masses and Yukawa couplings given by
\bea
\label{delta2}
\Delta_2^d &\approx& {f(c_q)^3} \frac{Y_*^2 v_4^2}{M_{KK}^2}
f(-c_d)v_4 Y_*
\eea
Similarly, the correction to the kinetic term of $d_R^{SM}$ gives the
contribution
\bea
\label{delta22}
{\Delta_2^d}' &\approx& f(c_q) \frac{Y_*^2 v_4^2}{M_{KK}^2}
f(-c_d)^3v_4 Y_*
\eea
Adding all terms together, we find the total fermion mass-Yukawa shift
\bea
\label{dapprox}
\Delta^d=\Delta^d_1+\Delta^d_2+{\Delta^d_2}'& \approx& f(c_q) \frac{Y_*^2 v_4^2}{M_{KK}^2}
f(-c_d)v_4 Y_*\left[2 + {f(c_q)^2} + f(-c_d)^2\right]
\eea

If we extend to the case of three generations, we can see that this shift
between SM fermion masses and Yukawa couplings produces a
misalignment in flavor space between these.
This misalignment will lead to flavor violating Higgs couplings
once the fermion mass matrix is diagonalized.

For the first two generation quarks, we need $f(c_q), f(-c_d) \ll 1$
to reproduce their small masses. Therefore, for these first two
generations, the shift coming from the correction to kinetic terms
(Fig. \ref{insertion1}) is negligible and the correction coming from
the diagrams in Fig. \ref{insertion} will dominate.
However, for the third generation, all effects are
comparable. It is interesting to point out that the expression
(Eq. \ref{dapprox}) (valid for one generation) is always positive,
which leads to a reduction in the 4d effective Yukawa couplings
compared to the SM ones.

\subsection{Brane Higgs subtlety}

Finally, we must mention that there is a subtlety in the case of an
exactly brane localized Higgs. As pointed out in \cite{BlankedeltaF2,Gori}, since the
wavefunctions of $q_R^{KK}$ and $d_L^{KK}$ vanish at TeV brane (due to
Dirichlet boundary conditions), their couplings to a brane localized Higgs should also
vanish. This means that the second diagram in Fig. \ref{insertion}
should give no contribution to the fermion mass-Yukawa shift (or at
best a highly suppressed one). We would then expect to be left with
only the correction coming from the kinetic term (Fig.
\ref{insertion1}), which as stated above is negligible for light
quarks. We observe, however, that upon EWSB, the wavefunctions
$q_R^{KK}$ and $d_L^{KK}$ become discontinuous at the brane location
\cite{Csaki:2003sh}, with the jump of the wavefunctions being
proportional to the brane Higgs vev $v_4$. This discontinuity
requires some sort of regularization of the brane location, meaning
that the couplings of $q_R^{KK}$ and $d_L^{KK}$ with the brane Higgs
would be infinitesimally small, but non-zero. But we note that in
the second diagram of Fig. \ref{insertion}, one has to sum over
infinite KK modes and even though each KK mode will give an
infinitesimally small contribution, the sum of infinite terms can
lead to a finite (non-zero) result (and as it turns out, this is
what happens, as shown explicitly in Appendix \ref{appendixIII} for
this mass insertion approximation).

This brane Higgs issue is avoided in \cite{NeubertRS}
because the authors did not include in their brane action any
operator of the type $H Q_R D_L$. By avoiding these, the
contribution to the shift $\Delta^d$ coming from the diagrams of
Fig. \ref{insertion} is simply not present (except for highly
suppressed corrections of order $\frac{v_4^2m_f^2}{M_{KK}^4}$ which
are safe to ignore).

We will address thoroughly this issue in the next two Sections and
again in Appendix \ref{appendixIII}, since we do find that the flavor
misalignment produced by the diagrams of Fig. \ref{insertion} is
large and of the same order for both bulk Higgs and brane Higgs
scenarios.

%%%%%%%%%%%%%%%%%%%%%%%%%%%%%%%%%%%%%%%%%%%%%%%%%%%%%%%%%%%%%%%%%%%%%%%%%%%%%%%%%%%%%%%%%%%%%%%%%%%%%
\section{5D calculation: Bulk Higgs Scenario}\label{bulk}

In this section we perform a 5D calculation in order to evaluate
more precisely the shift between Yukawa couplings and masses of SM
fermions. We start by working with a single fermion generation for
clarity but will later extend our results to the three generations case.

To proceed, we will need to solve for the wavefunctions of SM
fermions along the fifth dimension in the bulk Higgs
\cite{bulkhiggs,bulkhiggs1} scenario. This corresponds to  including
the contribution of all KK modes of the mass insertion
approximation, and not just the lightest ones. As we will see, the
most important shift does not go away as we push the Higgs profile
towards the IR brane. In the bulk Higgs scenario, the Higgs comes
from a 5D scalar with the following action \cite{bulkhiggs}
\begin{equation}
{\cal L}_{\text{Higgs}} = \int dz d^4x \left(\frac{R}{z}\right)^3
\left[ Tr |{\cal{D}}_M H|^2 - \frac{\mu^2}{z^2} Tr|H|^2  \right] -
V_{UV}(H)\delta(z-R) - V_{IR}(H)\delta (z-R')
\end{equation}
where $\mu$ is the 5D mass for Higgs in unit of $k$. The boundary
potentials $V_{UV}(H)$ and $V_{IR}(H)$ give the boundary conditions
for the Higgs wavefunction. We can choose these boundary conditions
such that the profile of the Higgs vev takes the simple form
\begin{equation}
v(z)=V(\beta)\ z^{2+\beta} \label{vz}
\end{equation}
where $\beta = \sqrt{4+\mu^2}$ and
\begin{eqnarray}\label{vevprof}
V(\beta) = \sqrt{\frac{2(1+\beta)}{R^3 (1-
(R'/R)^{2+2\beta})}}\frac{v_4}{(R')^{1+\beta}}
\end{eqnarray}
where $v_4$ is the SM Higgs vev. This nontrivial vev $v(z)$ is localized
towards the IR brane solving the Planck-weak hierarchy problem.
Nevertheless we will treat the brane Higgs case separately
later to review possible subtleties inherent to its localization by a
Dirac delta function.

After writing the 5D fermions in two component notation,
$Q=\left(\begin{array}{c}\calQ_L\\ {\calQ}_R\end{array}\right)$
and
$D=\left(\begin{array}{c}\calD_L\\ {\calD}_R\end{array}\right)$,
we perform a ``mixed'' KK decomposition as
\bea
\calQ_L(x,z) & =&   q_L(z)\, Q_L (x)  + ...\label{kkdecomp1}\\
{\calQ}_R(x,z)& =& q_R(z)\, {D}_R(x)+...\label{kkdecomp2}\\
\calD_L(x,z) & =&   d_L(z)\,  Q_L(x) +...\label{kkdecomp3}\\
{\calD}_R(x,z)& =&  d_R (z)\, {D}_R (x) +...\label{kkdecomp4}
\eea
where $Q_L(x),\ D_R(x)$ correspond to the light 4D SM fermions and the
$...$ include the rest of heavy KK fermion fields.
$q_{L,R}(z),\ d_{L,R}(z)$ are the corresponding profiles of the 4D SM
fermions  $Q_L(x)$ and $D_R(x)$ which verify the Dirac equation
\bea
-i \bar{\sigma}^{\mu} \partial_\mu Q_L(x) + m_d\, {D}_R(x) &=& 0, \\
-i \sigma^{\mu} \partial_\mu {D}_R(x) + m^*_d\, Q_L(x)& =& 0,
\eea
with $m_d$ being the 4D SM down-type quark mass (the analysis can be
carried out for up-type quarks in similar fashion).

The four profiles $q_{L,R}(z)$ and $d_{L,R}(z)$ must verify the
coupled equations coming from the equations of motion.
\bea
&& -m_d\ q_L - q'_R + {c_{q}+2\over z} q_R + \left({R\over  z}\right)v(z) Y_d\ d_R=0\label{qr}\\
&& -m^*_d\ q_R + q'_L + {c_{q}-2\over z} q_L + \left({R\over z}\right)v(z) Y_d\ d_L=0\label{ql}\\
&& -m_d\ d_L - d'_R + {c_{d}+2\over z} d_R + \left({R\over z}\right)v(z) Y^*_d\ q_R=0\label{dr}\\
&& -m^*_d\ d_R + d'_L + {c_{d}-2\over z} d_L + \left({R\over
z}\right) v(z) Y^*_d\ q_L=0\label{dl} \label{eqmot} \eea where the
$'$ denotes derivative with respect to the extra coordinate $z$ and
$[Y_d]=-1/2$ is 5D Yukawa coupling. Even if one knows the analytical
form of the nontrivial Higgs vev $v(z)$, solving analytically this
system of equations might still be quite hard. Nevertheless
it is  simple to find the misalignment between Higgs Yukawa
couplings and fermion masses based on the previous equations. To
proceed, let us first multiply Eq.~(\ref{qr}) by $q^*_L(z)$ and the
conjugate of Eq.~(\ref{ql}) by $q_R(z)$, and then subtract them.
One obtains \bea m_d(|q_L|^2-|q_R|^2)+ z^4\left({q^*_Lq_R\over
z^4}\right)'\!-\! \left({R\over z}\right) v(z)(Y_d d_Rq^*_L- Y^*_d
q_Rd^*_L)=0\ \ \eea We can now multiply by $R^4\over z^4$ and
integrate the whole expression between $z=R$ and $z=R'$ and obtain
\bea && R^4\int^{R'}_R dz \left(\frac{m_d}{z^4}
(|q_L|^2-|q_R|^2)-{Rv(z)\over z^5}(Y_d d_Rq^*_L-Y^*_d
q_Rd^*_L)\right) + \left(q^*_Lq_R {R^4\over z^4}\right)\Big|^{R'}_R\
=\ 0 \label{misal1} \eea The boundary conditions for the profile
$q_R(z)$ are chosen to be Dirichlet at both boundaries, i.e.
$q_R(R)=q_R(R')=0$, which means that the last term of
Eq.~(\ref{misal1}) identically vanishes. Moreover, canonical
normalization of the SM d-quark imposes the extra constraint \bea
R^4\int^{R'}_R\frac{dz}{z^4}(|q_L|^2+|d_L|^2) \ = \ 1. \eea We can
therefore rewrite Eq.~(\ref{misal1}) as \bea m_d=R^4\int^{R'}_R\!\!
dz \left(\frac{m_d}{z^4} (|d_L|^2+|q_R|^2) + {Rv(z)\over z^5}(Y_d
d_Rq^*_L\!-\!Y^*_dq_Rd^*_L)\right) \ \ \eea Note that this identity
is exact, but also that each profile $q_{R,L}(z)$ and $d_{R,L}(z)$
depend on the mass $m_d$. In the zero mode approximation, the
profiles with Dirichlet boundary conditions, $q^0_R(z)$ and
$d^0_L(z)$ vanish, and the identity can be expressed as \bea
m_d\simeq m^0_d=R^5\int^{R'}_R\!\! dz {v(z)\over z^5}Y_d
d^0_Rq^{0*}_L \ \ \eea which agrees with the intuition that fermion
mass is mostly generated by the 5D Yukawa couplings between the 5D
Higgs and the zero mode fermion profiles. From the action in
Eq.~(\ref{fermionaction}) we also extract the 4D Yukawa coupling of
the Higgs field (the lightest KK mode of the 5D Higgs) and the SM
down type quark. \bea y^d_4 = R^5\int^{R'}_R dz {h(z)\over z^5}(Y_d
d_Rq_L^*+ Y^*_dq_Rd_L^*) \eea where $h(z)$ is the profile of the
physical Higgs field. It is easy to show that the Higgs vev solution
$v(z)$ is related to the profile of the physical light Higgs $h(z)$
(lightest KK mode) by \bea h(z)=\frac{v(z)}{v_4} \left(1+{\cal
O}\left(\frac{ m_h^2 {z}^2}{1+\beta}\right) \right) \eea so for a
light enough Higgs field both profiles $h(z)$ and $v(z)$ are
proportional to each other. For a moderately heavy physical Higgs,
there will be a misalignment between the profiles of the Higgs vev
and the physical Higgs, leading to a misalignment between fermion
masses and Yukawa couplings. However, this effect can actually be
decoupled if the Higgs is pushed towards the IR brane (by increasing
the parameter $\beta$). In this case, the Higgs vev profile will be
more and more aligned with that of the physical Higgs, so that they
become identical in the brane Higgs limit. This source of Higgs
flavor violating couplings will be controlled by the parameter
$\frac{1}{\beta+1}$ and for the sake of clarity we will ignore its
effects in the rest of the paper because, as we discuss in Appendix
\ref{appendixII}, they are numerically small and can be decoupled by
pushing the Higgs towards the IR brane.

We can then compute the shift $\Delta^d=m_d - v_4\ y^d_4$ between
the fermion mass $m_d$ and the Yukawa coupling $y^d_4$ as
\bea
\Delta^d&=&R^4\int^{R'}_R\!\! dz
\left(\frac{m_d}{z^4} (|d_L|^2+|q_R|^2) -2Y^*_d {R v(z)\over z^5}
q_Rd_L^*\right). \
\label{Delta}
\eea This identity  shows that the
shift has to be relatively small since it vanishes in the zero mode
approximation.

To proceed further, we will use a perturbative approach such that we
assume that $(v_4R')\ll 1$ where $v_4$ is the SM Higgs vev. Thus, once
we know the analytical form of the vev profile $v(z)$ (see
Eq.~(\ref{vz})) we can solve perturbatively the system of coupled
equations (\ref{qr}-\ref{dl})\footnote{It would be interesting to
  use this perturbative technique in the context of fermion flavor in soft-wall
  scenarios \cite{Batell:2008me,Delgado:2009xb,Aybat:2009mk} given
  that the setup is quite similar; we will leave this analysis for future studies.}.

We find
\bea
q_L(z)&=&Q_L\ z^{2-c_{q}}\left[1+\ {\cal O}(v_4^2{R'}^2)\right]\label{qlsol}\\
d_R(z)&=&D_R\ z^{2+c_{d}}\left[1+\ {\cal O}(v^2_4{R'}^2) \right]\eea 
and 
\bea
\hspace{-.3cm}q_R(z)&=&\left[m_d\ Q_L \left({ R^{1-2c_{q}}\over 1-2c_{q}} z^{2+c_{q}} -
{1 \over 1-2c_{q}}\ z^{3-c_{q}}\right) +\ {Y_d } {R V(\beta)\over
(2+\beta -c_q+c_{d})} D_R\ z^{4+\beta+c_d}\ \right ]
\left[1+\ {\cal O} (v_4^2{R'}^2)\right] \ \ \ \ \label{qrsol} \\
\hspace{-.3cm}d_L(z)&=&\left[m^*_d\ D_R \left(-{ R^{1+2c_{d}}\over 1+2c_{d}} z^{2-c_{d}}
+ {1 \over 1+2c_{d}}\ z^{3+c_{d}}\right) -\ {Y^*_d } {R V(\beta)\over
(2+\beta -c_q+c_{d})} Q_L\ z^{4+\beta-c_q}\right]\left[1+\ {\cal O} (v_4^2{R'}^2)\right]
\label{dlsol}
\eea with the constants $Q_L$ and $D_R$ fixed by canonical
normalization of the kinetic terms giving
\bea
Q_L&=&\sqrt{\frac{1-2c_{q}}{\epsilon^{2c_{q}-1}-1}} R^{c_{q}-5/2}\\
D_R&=&\sqrt{\frac{1+2c_{d}}{\epsilon^{-2c_{d}-1}-1}} R^{-c_{d}-5/2}
\eea
Equipped with the solutions from Eqs.~(\ref{qlsol}) to (\ref{dlsol})
one can evaluate perturbatively the shift $\Delta^d$ defined in
Eq.~(\ref{Delta}). For simplicity, we present here the results for
UV localized fermions ($c_q>0.5,c_d<-0.5$). The general results for
both UV and IR localized fermions are presented in Appendix
\ref{appendixI}. We find that the main contribution to the shift
coming from the last term in Eq.~(\ref{Delta}) can be written as
\bea
\label{y3}
\Delta^d_{1} &=& 2 |m_d|^2 m_d
{R'}^2\left[\frac{(2+\beta+c_d-c_q)}{(6+3\beta+c_d-c_q)} - 2
\frac{(2+\beta + c_d - c_q)}{(2\beta + 4)} + \frac{(2+\beta + c_d -
c_q)}{(2+\beta + c_q - c_d)}\right]\frac{1}{f(c_q)^2 f(-c_d)^2}
\eea
This result corresponds to the one we estimated earlier by using the insertion
approximation (see Eq.~(\ref{delta1})).

The first term in Eq.~(\ref{Delta}) gives a subleading contribution to the shift
\bea
\label{y4}
\Delta^d_{2}= m_d |m_d|^2 R'^2 \left[ \frac{1}{f(c_q)^2}\left( \frac{2
  c_q -1}{2c_q+1} + \frac{1}{5 + 2\beta + 2c_d} - \frac{1}{3+c_q+c_d
  + \beta}\right) + ( c_{q,d} \to - c_{d,q})  \right]
\label{kinetic}
\eea
which corresponds to the one coming from the kinetic correction
using the insertion approximation (Eq. \ref{delta2} and \ref{delta22}).

Even if the fermion mass $m_d$ is small, the large warp factor
$\frac{1}{f(c_q)^2 f(-c_d)^2}\approx \epsilon^{2-2c_q+2c_d}$ will
overcome most of the suppression, rendering the shift to be of the
order $ \Delta^d\sim m_d  v_4^2 {R'}^2 $.
The shift is generally on the percent level with respect to fermion
masses, but a misalignment of this order in the Higgs Yukawa
couplings should introduce strong constraints due to FCNC's.

\subsection{Pushing the Higgs from the bulk to the brane}

Note that in the $\beta \rightarrow \infty$ limit, the profile of
the Higgs vev tends to become brane localized, as well as the light
physical Higgs and the rest of Higgs KK modes. In this limit, the
shift $\Delta^d_1$ produced between the fermion mass and the Yukawa
coupling, coming from the diagrams of Fig. \ref{insertion}, reduces
to \bea \Delta^d_{1}=\frac{2}{3} |m_d|^2 m_d{R'}^2\frac{1}{f(c_q)^2
f(-c_d)^2}, \eea and in particular we see that the effect does not
decouple (i.e. it is non-zero). The fact that the expected
misalignment is more or less independent on the localization of the
Higgs is one of our main results since the bounds and predictions
that we will extract can then be considered a general feature of RS
models with fields in the bulk (and a Higgs scalar localized near or
at IR brane)\footnote{An interesting exception to these results
  in the Higgs sector, proposed in \cite{Agashe:2009di}, would be to
  eliminate the Higgs as a fundamental scalar and consider the fifth
  component of a gauge field as playing the Higgs role in EWSB.}.
The shift $\Delta^d_2$ coming from the corrections to the fermion
kinetic terms (Fig. \ref{insertion1}) becomes in the $\beta
\rightarrow \infty$ limit \bea \Delta^d_{2}= m_d |m_d|^2 R'^2 \left[
\frac{1}{f(c_q)^2}\left( \frac{2
  c_q -1}{2c_q+1} \right) +
 \frac{1}{f(-c_d)^2}\left( \frac{2c_d +1}{2c_d-1} \right)
%+ ( c_{q,d} \to - c_{d,q})
\right],
\eea
in agreement with the results found in \cite{NeubertRS} (for a brane
Higgs scenario).

Maybe it can be useful to discuss the validity of the $\beta\rightarrow
\infty$ limit starting from a bulk Higgs scenario.
Let's first look at the mass spectrum in this case. The Higgs profile is given by
Eq.~(\ref{higgsprofile}) and to find its mass eigenvalues one has to
satisfy the appropriate boundary conditions at the IR brane \cite{bulkhiggs}
\begin{eqnarray}
\left. \partial_z h +\left(\frac{R'}{R}\right)m_{\text{TeV}} h\right
|_{R'}=0.
\end{eqnarray}
This will lead to one light mode (i.e. SM Higgs) and a tower of heavy
modes with masses proportional to $\sim \beta/R'$, and so in the 
$\beta\rightarrow \infty$ limit all the KK Higgs excitations are
decoupled from the low energy spectrum. This means that in this
limit we can treat Higgs field as an effective four dimensional
field, and thus it corresponds to the brane Higgs scenario.
As mentioned earlier (and in Appendix \ref{appendixII}), the misalignment
caused by a difference in profiles between the Higgs physical field
and its vev (and which we have neglected) will also disappear, as
one can interpret that specific misalignment as a result of the mixing between SM Higgs and the
heavy Higgs KK modes, which is controlled by $\frac{1}{\beta}\sim
\frac{1}{M^{\text{Higgs}}_{\text{KK}} R'}$.

Let us now look on the couplings of fermions to the Higgs in this limit.
For the zero  modes we will get:
\begin{eqnarray}
y^{SM}_d=\frac{\sqrt{2(1+\beta)}}{(2-c_q+c_d+\beta)}\frac{Y_d}{\sqrt{R}} f(c_q)f(-c_d)
\end{eqnarray}
where $[y^{SM}_d]=0,[Y_d]=-1/2$; similarly one can look at the
couplings of two KK fermions to the Higgs and in this case one finds
its dependence to be $\sim
\frac{1}{\sqrt{\beta}}\frac{Y_d}{\sqrt{R}}$. Naively both couplings
do vanish in the $\beta\rightarrow\infty$ limit. But if the 5D
couplings $Y_d$ scale as $\sqrt{\beta}$ then these couplings will
have a finite limit given by the usual brane Higgs results. One can
argue whether we can scale the 5D Yukawas as $\sqrt{\beta}$ because
such large Yukawas should violate perturbativity of the theory, but
as was shown above the couplings of the Higgs to the KK fermions are
still $O(1)$. One can see that  only the KK excitations of the Higgs
will have couplings with KK fermions $\sim Y_d R^{-1/2} \propto
O(\sqrt{\beta})$, but their masses are $O(\frac{\beta}{R'})$ and
they are completely decoupled from the spectrum. So we conclude this
discussion by stressing that it is consistent to consider the limit
$\beta\rightarrow \infty$ with $Y_d \propto \sqrt{\beta}$ and it
coincides with the usual brane Higgs scenario.

%%%%%%%%%%%%%%%%%%%%%%%%%%%%%%%%%%%%%%%%%%%%%%%%%%%%%%%%%%%%%%%%%%%%%%%%%%%%%%%%%%%%%%%%%%%%%%%%%%%%%
\section{5D calculation: Brane Higgs Scenario}\label{brane}

We argued in Section \ref{diagrammatic} that one might expect that
the major contribution to the misalignment $\Delta^d_{1}$ vanishes
in the brane Higgs case since the odd KK modes $q_R^{KK}$,
$d_L^{KK}$ have vanishing wavefunctions on the IR brane. We also
briefly mentioned that in the mass insertion approximation, one
actually might need to sum the infinite tower of fermion KK modes to
obtain a non-vanishing contribution (see Appendix \ref{appendixIII}
for details). However, without invoking that explanation, we just
saw that in the $\beta \to \infty$ limit, $\Delta^d_{1}$ approaches
a nonzero value of same numerical order as the $\beta=finite$ case.
Since the $\beta \to \infty$ limit of bulk Higgs corresponds to a
brane localized Higgs, there seems to be a counter-intuitive
subtlety. In this section we try to address and resolve this point
in a more precise way, by performing the 5D calculation of the shift
$\Delta^d_{1}$ for the specific scenario of a brane Higgs.

For brane Higgs, we can write the Yukawa couplings in the Lagrangian as
\begin{equation}
S_{\text{brane}} = \int d^4x dz\, \delta(z- R') \left(\frac{R}{z}
\right)^4 H \left( {Y^{5D}_1} R \bar{{\cal Q}}_L {\cal D}_R +
{Y^{5D}_2} R\bar{{\cal Q}}_R {\cal D}_L + \text{h.c.} \right)
\label{braneaction}
\end{equation}
Here we choose the convention with $\text{dim}[ Y^{5D}_{1,2}]=0$.
Note that compared to the bulk Higgs case, the Yukawa couplings
$Y^{5D}_1$ an $Y^{5D}_2$ are independent and both $\sim O(1)$.
However, they should be of the same order due to the philosophy of
flavor anarchy and naturalness. We can do KK decomposition as
before, then the equations satisfied by the wavefunctions are
\begin{eqnarray}
\label{eom}
-m_d q_L - \partial_z q_R + \frac{c_q+2}{z}q_R + v_4\delta(z-R')Y^{5D}_1 R' d_R = 0\\
-m^*_d q_R + \partial_z q_L + \frac{c_q-2}{z}q_L + v_4\delta(z-R')Y^{5D}_2 R' d_L = 0 \\
-m_d d_L - \partial_z d_R + \frac{c_u+2}{z}d_R + v_4\delta(z-R')Y^{5D*}_2 R' q_R = 0\\
-m^*_d d_R + \partial_z d_L + \frac{c_u-2}{z}d_L +
v_4\delta(z-R')Y^{5D*}_1 R' q_L = 0
\end{eqnarray}
Notice that the odd wavefunctions $q_R$ and $d_L$ vanish at the IR
brane. But the delta functions in equations above give a jump for
$q_R$ and $d_L$ at the IR brane, which makes their values at IR brane
ambiguous \cite{Csaki:2003sh}. To remove this ambiguity, we ``regularize'' the delta
in the following way
\begin{equation}
\label{regulator} \delta(z-R') = \lim_{\varepsilon \rightarrow 0}
\begin{cases}
\frac{1}{\varepsilon}, ~~R'-\varepsilon<z<R'\\
0, ~~z<R'-\varepsilon.
\end{cases}
\end{equation}
This regularization is in a way similar to treating the Higgs as a
bulk field and then taking the limit $\beta\to \infty$, although
without apparent divergences coming from taking $\beta$ to be
large. In any case one could also perform other regularization methods
to remove the wavefunction ambiguities at the IR brane\footnote{For
  example, we could have chosen instead to move the delta function
  location from $R'$ to $(R'-\varepsilon)$, and enforce the usual
  boundary conditions on the fields at $z=R'$. Then, at the very end,
  we would take the limit $\varepsilon\to 0$ \cite{Csaki:2003sh}. In that
  case we find
\bea d_L(z),q_R(z) &\propto& v_4 Y^{5D}_1\
\theta(z-R'+\varepsilon)\qquad {\rm for}\ \ R'-2\varepsilon<z<R',
\eea 
where we have used the step function $\theta(x)=1$ for $x<1$
and $\theta(x)=0$ for $x>0$. Inserting this into Eq.~(\ref{Delta})
we obtain the same misalignment as in Eq.~(\ref{Deltabrane1}),
namely 
\bea 
\Delta^d_1 &\propto& 2 (v_4R')^3 (Y^{5D}_1)^2Y^{5D*}_2\
\int_{R'-2\varepsilon}^{R'} dz\ \delta(z-R'+\varepsilon)\
\left[\theta(z-R'+\varepsilon)\right]^2\ \ \ \ \propto\ \
\frac{2}{3} (v_4R')^3 (Y_1^{5D})^2Y_2^{5D*} .\non 
\eea }.

Now we can easily impose Dirichlet boundary conditions for the $q_R,d_L$ profiles at IR brane
\begin{eqnarray}
q_R(R')=d_L(R')=0
\end{eqnarray}
Integrating equations of motion (Eq. \ref{eom}) from
($R'-\varepsilon<z<R'$) will lead to
\begin{eqnarray}
q_R(R')-q_R(R'-\varepsilon) &=& v_4Y^{5D}_1 R' d_R(R')\\
d_L(R')- d_L(R'-\varepsilon) &=& -v_4Y^{5D*}_1 R' q_L(R')
\end{eqnarray}
For the rectangular potential profiles $q_R,d_L$ will drop to zero
linearly in the region $R'-\varepsilon<z<R'$, so the profiles near the IR
brane can be approximated by
\begin{eqnarray}
\label{oddwfbrane}
q_R(z) = v_4Y^{5D}_1 R' d_R(R')\left( \frac{z-R'}{\varepsilon}\right)\qquad
{\rm for} \ \ R'-\varepsilon<z<R',\\
d_L(z) = -v_4Y^{5D*}_1 R' q_L(R')\left(
\frac{z-R'}{\varepsilon}\right)\qquad{\rm for} \ \ R'-\varepsilon<z<R'.
\end{eqnarray}
From our previous discussion, the main contribution to the
misalignment between SM fermion masses and Yukawa couplings come
from the second term of Eq.( \ref{Delta}), so plugging in the odd
wavefunctions from Eq.(\ref{oddwfbrane}), we get
\begin{eqnarray}
\Delta^d_{1}&=&2 (Y^{5D}_2)^* (Y^{5D}_1)^2 R'^3 v_4^3 d_R(R')
q^*_L(R')\left(\frac{R}{R'}\right)^4 \int_{R'-\varepsilon}^{R'} dz
\frac{1}{\varepsilon}\left( \frac{z-R'}{\varepsilon}\right)^2\nonumber\\
&=&\frac{2}{3}(Y^{5D}_2)^* (Y^{5D}_1)^2 R'^3 v_4^3 d_R(R')
q^*_L(R')\left(\frac{R}{R'}\right)^4\label{Deltabrane1}
\end{eqnarray}
On the other hand, to leading order in Higgs vev, the SM fermion
mass is given by
\begin{equation}
m_d \approx \left(\frac{R}{R'}\right)^4 v_4 Y^{5D}_1R' q^*_L(R')
d_R(R')
\end{equation}
Therefore, the misalignment can be expressed as
\begin{equation}
\label{deltabrane}
\Delta^d_{_1}=\frac{2}{3} m_d Y^{5D}_1  (Y^{5D}_2)^* v_4^2 R'^2 =
\frac{2}{3} |m_d|^2 m_d R'^2
\left(\frac{Y_2^{5D}}{Y_1^{5D}}\right)^* \frac{1}{f(c_q)^2f(-c_d)^2}
\end{equation}
As advertised before, this result agrees with the one obtained in the previous section for
the bulk Higgs scenario, once we take $\beta \rightarrow \infty$
(Eq. \ref{y3}). We again stress that this result shows that upon
careful derivation, the misalignment obtained does not vanish in the
particular case of a Brane localized Higgs.
The main difference though, is the appearance of the independent couplings
$Y_2^{5D}$, which in the bulk Higgs case are forced to be equal to
$Y^{5D}_1$ by 5D general covariance. These couplings $Y^{5D}_2$ are not necessary
for generating fermion masses, and so it is technically possible to set
their values as small as necessary to suppress the obtained
misalignment. Nevertheless this seems to go against the main
philosophy of our approach which is to assume the value of all
dimensionless 5D parameters of order one.

Again, the fact that $\Delta^d_{1}$ is non zero in the brane Higgs
case is hard to understand 
in the mass insertion approximation since the contribution from each
KK fermion (see Fig. \ref{insertion}) seems to be vanishing. In
Appendix \ref{appendixIII} we show that to resolve this point we
need to sum up all the KK modes of the mass insertion approximation,
as already mentioned before.

The subleading contribution to the misalignment between SM fermion
masses and Yukawa coupling can be calculated in a similar way as in
the previous section, and the result is (for UV localized fermions)
\begin{eqnarray}
\Delta^d_{2} &=& m_d |Y^{5D}_1|^2 v_4^2 R'^2 \left[ f(-c_d)^2
\frac{2c_q - 1}{2c_q+1} + (c_{q,d} \to - c_{d,q}) \right]\\ &=& m_d
|m_d|^2 R'^2\left[ \frac{1}{f(c_q)^2}\left(\frac{2c_q -1}{2c_q
+1}\right) + (c_{q,d}\to -c_{d,q}) \right]
\end{eqnarray}
We can see that for the first two generations, we have $\Delta^d_{2}
\ll \Delta^d_{1}$, and it agrees with Eq. (\ref{y4}) in the $\beta
\to \infty$ limit. The result for both UV and IR localized fermions
is given by
\begin{eqnarray}
\Delta^d_{2} = m_d |m_d|^2 R'^2 \left[K(c_q) + K(-c_d)\right]
\end{eqnarray}
with \begin{equation} K(c) \equiv \frac{1}{1-2c}\left[
-\frac{1}{\epsilon^{2c-1}-1}+ \frac{\epsilon^{2c-1}-\epsilon^2}
{(\epsilon^{2c-1}-1)(3-2c)}+\frac{\epsilon^{1-2c}-\epsilon^2}
{(1+2c)(\epsilon^{2c-1}-1)}\right].
\end{equation}
One can see that $\Delta^d_{1}$ and $\Delta^d_{2}$ can be of the same
order only for IR localized fermions.

%%%%%%%%%%%%%%%%%%%%%%%%%%%%%%%%%%%%%%%%%%%%%%%%%%%%%%%%%%%%%%%%%%%%%%%%%%%%%%%%%%%%%%%%%%%%%%%%%%%%%
\section{Generalizing to three Generations}

We can generalize the calculations presented in the sections
\ref{bulk} and \ref{brane} to 3 generations. For simplicity we perform the
analysis in the brane Higgs scenario here. %(i.e. $\beta\rightarrow \infty$).
To leading order in Yukawa, the SM fermion mass matrix is
\begin{equation}
\hat{m}^d_{\alpha \beta} = [\hat{F}_q \hat{Y}_{1}^{5D} \hat{F}_d]_{\alpha \beta} {v_4}
\end{equation}
where $\hat{}$ means a $3 \times 3$ matrix in flavor space and
$\hat{F}_{q,d} = \text{diag}[f({c_{q_i}, c_{d_i}})]$. Using the same
technique as before, we can easily show that the misalignment
between fermion mass and Yukawa coupling matrix is $\hat{\Delta}^d =
\hat{\Delta}^d_1 + \hat{\Delta}^d_2$, with

\bea
\label{threegenresult11}
\hat{\Delta}^d_{1,\alpha \beta}&=&\frac{2}{3}\left[ \hat{F}_q
\hat{Y}^{5D}_1 (\hat{Y}^{5D}_2)^\dagger \hat{Y}^{5D}_1 \hat{F}_d
\right]_{\alpha \beta}\left({v_4^3 R'^2}\right)\\
&=& \frac{2}{3}\left[\hat{m}^{d} \frac{1}{\hat{F}_d}
  (\hat{Y}^{5D}_2)^\dagger \frac{1}{\hat{F}_q} \hat{m}^{d}
  \right]_{\alpha \beta}\left({v_4^3 R'^2}\right)
\label{threegenresult12}
\eea
and
\begin{equation}
\label{threegenresult2}
\hat{\Delta}^d_{2,\alpha \beta} = \left[\hat{m}^d \left(
\hat{m}^{d\dagger} \hat{K}(c_q) + \hat{K}(-c_d) \hat{m}^{d\dagger}
\right)\hat{m}^d\right]_{\alpha \beta} R'^2
\end{equation}
The subdominant contribution here (Eq. \ref{threegenresult2}) agrees
with the result found in \cite{NeubertRS}. The crucial observation
is that $\hat{m}^d_{\alpha \beta} $ and $\hat{\Delta}^d_{\alpha
\beta}$ are generally not aligned in flavor space. Thus when we
diagonalize the quark mass matrix with a bi-unitary transformation
$\hat{m}^d \rightarrow O^\dagger_{d_L} \hat{m}^d O_{d_R}$, the
Yukawa couplings will not be diagonal. To be more specific, in
models of flavor anarchy,
we have
\begin{equation}\label{fratio}
(O_{d_L, d_R})_{\alpha \beta} \sim
\frac{F_{q_\alpha,d_\alpha}}{F_{q_\beta,d_\beta}} \qquad \text{for}
\quad {\alpha < \beta}
\end{equation}
Then the off-diagonal Yukawa coupling will be (dominated by Eq.
(\ref{threegenresult11}))
\begin{eqnarray}\label{offcoup}
\hat{Y}^{\text{off}}_{\alpha \beta}& = & -(O^\dagger_{d_L}
\hat{\Delta}^d O_{d_R})_{\alpha \beta}\frac{1}{v_4}\\ \nonumber &\sim &
\frac{2}{3} F_{q_\alpha}\bar{Y}^3 F_{d_\beta} v_4^2 R'^2
\end{eqnarray}
where $\bar{Y}$ is the typical value of the dimensionless 5D Yukawa coupling.

%%%%%%%%%%%%%%%%%%%%%%%%%%%%%%%%%%%%%%%%%%%%%%%%%%%%%%%%%%%%%%%%%%%%%%%%%%%%%%%%%%%%%%%%%%%%%%%%%%%%%
\section{Estimates of Higgs FCNC in Flavor Anarchy}

In this section, we estimate the off-diagonal couplings of Higgs to
SM fermions (assuming again for simplicity a brane Higgs
scenario). And then we do a numerical scan over anarchical Yukawa
couplings to support our estimates. We first parametrize the
Higgs Yukawa couplings as
\begin{equation}
{\cal{L}}_{HFV} = a^d_{ij}\sqrt{\frac{m^d_i m^d_j}{v_4^2}} H
\bar{d}_L^i d_R^j + h.c. + (d \leftrightarrow u).
\end{equation}
We can use Eq. (\ref{fratio}) and (\ref{offcoup}) to estimate the
sizes of $a^{u,d}_{ij}$. For example, we have
\begin{eqnarray}
a^d_{12} &\sim& \frac{2}{3} F_{q_1} \bar{Y}^3 F_{d_2} v^2 R'^2  \sqrt{\frac{v_4^2}{m_s m_d}} \\
\nonumber &=&\frac{2}{3} \frac{F_{q_1}}{F_{q_2}} \bar{Y}^2 v_4 R'^2
F_{q_2} \bar{Y} v_4 F_{d_2} \sqrt{\frac{v_4^2}{m_s m_d}}\\ \nonumber
&\sim& \frac{2}{3} \lambda \bar{Y}^2 v_4^2 R'^2
\sqrt{\frac{m_s}{m_d}},
\end{eqnarray}
where $\lambda \approx 0.22$ is the Wolfenstein parameter, and we
used $F_{q_1}/F_{q_2} \sim (O_{d_{L}})_{12}\sim (V_{CKM})_{12}\sim
\lambda$. We can find the other $a^{u,d}_{ij}$'s in similar fashion.
Here we present our results from estimates:
\begin{eqnarray}
a^d_{ij} \sim \delta_{ij}  -\frac{2}{3}\bar{Y}^2 v_4^2
R'^2\left(\begin{array}{ccc} 1 &
\lambda\sqrt{\frac{m_s}{m_d}} & \lambda^3 \sqrt{\frac{m_b}{m_d}}\\
\frac{1}{\lambda}\sqrt{\frac{m_d}{m_s}} & 1 & \lambda^2
\sqrt{\frac{m_b}{m_s}} \\ \frac{1}{\lambda^3}\sqrt{\frac{m_d}{m_b}}
& \frac{1}{\lambda^2}\sqrt{\frac{m_s}{m_b}} & 1
\end{array} \right)\label{adest}
\end{eqnarray}
\begin{eqnarray}
a^u_{ij} \sim \delta_{ij} -\frac{2}{3}\bar{Y}^2 v_4^2
R'^2\left(\begin{array}{ccc} 1 &
\lambda\sqrt{\frac{m_c}{m_u}} & \lambda^3 \sqrt{\frac{m_t}{m_u}}\\
\frac{1}{\lambda}\sqrt{\frac{m_u}{m_c}} & 1 & \lambda^2
\sqrt{\frac{m_t}{m_c}} \\ \frac{1}{\lambda^3}\sqrt{\frac{m_u}{m_t}}
& \frac{1}{\lambda^2}\sqrt{\frac{m_c}{m_t}} & 1
\end{array} \right)\label{auest}
\end{eqnarray}
Note that the results we presented here are just estimates for the
size of $a^{u,d}_{ij}$, not their signs or phases. However, for the
third generation quarks, the corrections almost always suppress the
Yukawa couplings if $Y_1 = Y_2$ (which is natural in bulk Higgs
scenario) and are typically larger than the previous estimates. We
argue this point in the next subsection.

\subsection{Yukawa couplings of the third generation when $Y_1 = Y_2$}
\label{3rdgeneration}

We can obtain a better estimate on the typical size of the diagonal
entries of the Yukawa coupling matrices by going back to
Eq.~(\ref{threegenresult12}) and assume that $Y_1=Y_2$. Its form
simplifies further to
\bea
\hat{\Delta}^u_{1,\alpha\beta}&=& \frac{2}{3} R'^2 \left[\hat{m}^u
  \frac{1}{\hat{F}_u^2} (\hat{m}^u)^\dagger \frac{1}{\hat{F}_q^2}
  \hat{m}^u\right]_{\alpha\beta}
\eea
where we have written the misalignment in the up-sector.
Now we perform the bi-unitary rotation needed to go to the
physical fermion basis, and study the element (33) of the overall
Yukawa coupling, i.e.
\bea
a_{tt}-1&=&-\frac{2{R'}^2}{3m_t} \left[ O^\dagger_{u_L}
\hat{m}^u\frac{1}{\hat{F}_u^2}
\hat{m}^{u\dagger}\frac{1}{\hat{F}_q^2} \hat{m}^u O_{u_R}
\right]_{33} \non\\
&=&-\frac{2{R'}^2}{3m_t}
\left( m_u^{diag}\right)_{33}
\left(O^\dagger_{u_R}\frac{1}{\hat{F}_u^2} O_{u_R}\right)_{3j}
\left(m_u^{diag}\right)_{jj}
\left(O^\dagger_{u_L}\frac{1}{\hat{F}_q^2} O_{u_L}\right)_{j3}
\left(m_u^{diag}\right)_{33}
\eea
First let's look at the contribution to $a_{tt}$ when the ``$j$'' index
is equal to 3 (i.e. in the middle mass matrix $m_u^{diag}$ we have
$m_t$). In this case, there will be 9 terms, each proportional to
$-\frac{2{R'}^2 \bar{Y}^2 v_4^2}{3}  $, and it is important to
realize that every one of them will be real and negative, because
$(O^\dagger_{u_R}\frac{1}{\hat{F}_u^2} O_{u_R})_{33}\geq0$. When
$j=2$ $(m_c)$ there will be only 4 terms $\sim\frac{2{R'}^2
\bar{Y}^2 v_4^2}{3}$ but every one of them will have generically
a random complex phase (the 5 remaining terms are much
smaller).
 For $j=1$ $(m_u)$ there is only one term $\sim\frac{2{R'}^2
  \bar{Y}^2 v_4^2}{3}$ contributing, with the other 8 terms being
again suppressed. So at the end of the day the dominant contribution
to $a_{tt}$  will consist of 14 terms, 9 of which are negative and
the rest 5 have random complex phases. Generically each of these terms
are of the same size $\sim \frac{2{R'}^2 \bar{Y}^2 v^2}{3}$ so from
a statistical argument, $a_{tt}-1$ should receive a negative
contribution $\sim-9 \left( \frac{2{R'}^2 \bar{Y}^2 v^2}{3}\right)$.
This result is confirmed by the numerical scan presented below.

One can perform the same analysis for the element (22) of the Yukawa
matrix and realize that in this case the number of terms aligned
(contributing constructively) is 4, and for the (11) element there are
none. This means that the largest corrections are expected in the
third generation Yukawa couplings, with a suppressed correction in
second generation couplings and much more suppressed correction for
first generation couplings. This structure in the
corrections seems to be a result of the hierarchical structure of
the flavor anarchy setup.

Finally, we must remind the reader that it was crucial to take $Y_1 =Y_2$ (which is
required in the Bulk Higgs scenario) to obtain these predictions.
In the case $Y_1 \ne Y_2$, there will be no alignment of terms, and we
therefore generally expect smaller corrections to the third generation
Yukawa couplings.

\subsection{Validity of $\bar{Y}v_4 R'$ expansion}
We managed to solve the fermion equations by expanding them in the
parameter $(\bar{Y}^2 v_4^2 R'^2)$, and so our results can be trusted
as long as
\begin{eqnarray}
\bar{Y}\lesssim \frac{1}{v_4 R'}\ \ \ \left(\sim 9\,\, \text{for}
\,\,R'^{-1}=1500\hbox{GeV}\right)
\end{eqnarray}
but we have seen in the previous subsection that the corrections to
$htt$ and $hbb$ couplings do pick up an extra numerical factor of
$\sim 9$ in the expansion parameter $(\bar{Y}^2 v_4^2 R'^2)$. This
means that, at least for third generation fermions, our approximation
is valid only for
\begin{eqnarray}
{\bar{Y}}\lesssim \frac{1}{v_4 R'\sqrt{9}}\ \,\,\, \left( \sim 3\,\,
\text{for} \,\,R'^{-1}=1500\hbox{GeV}\right)
\end{eqnarray}
Generically for the case with $\bar{Y}\gtrsim 3 $ we will still have
a large misalignment between the Higgs couplings and fermion masses
but to be able to make valid predictions one would have to solve the
equations of motion (Eq. \ref{qr} to \ref{eqmot}) exactly or use a
different perturbative parameter. In the numerical analysis
presented below we performed a scan with $0.3<|Y^{5D}_{1,2}|<3$,
where our expansion is valid. We then also allowed for slightly
larger values of the Yukawas such that $1<|Y^{5D}_{1,2}|<4$. The
average size of the couplings is still below $3$, so for a KK scale
of $R'^{-1}=1500$ GeV or above, the results will still be precise
enough, although approaching the edge of perturbative convergence.

\subsection{Numerical Scan}

\begin{figure}[htp]
\vspace{-.cm} \center
    \includegraphics[height=5cm]{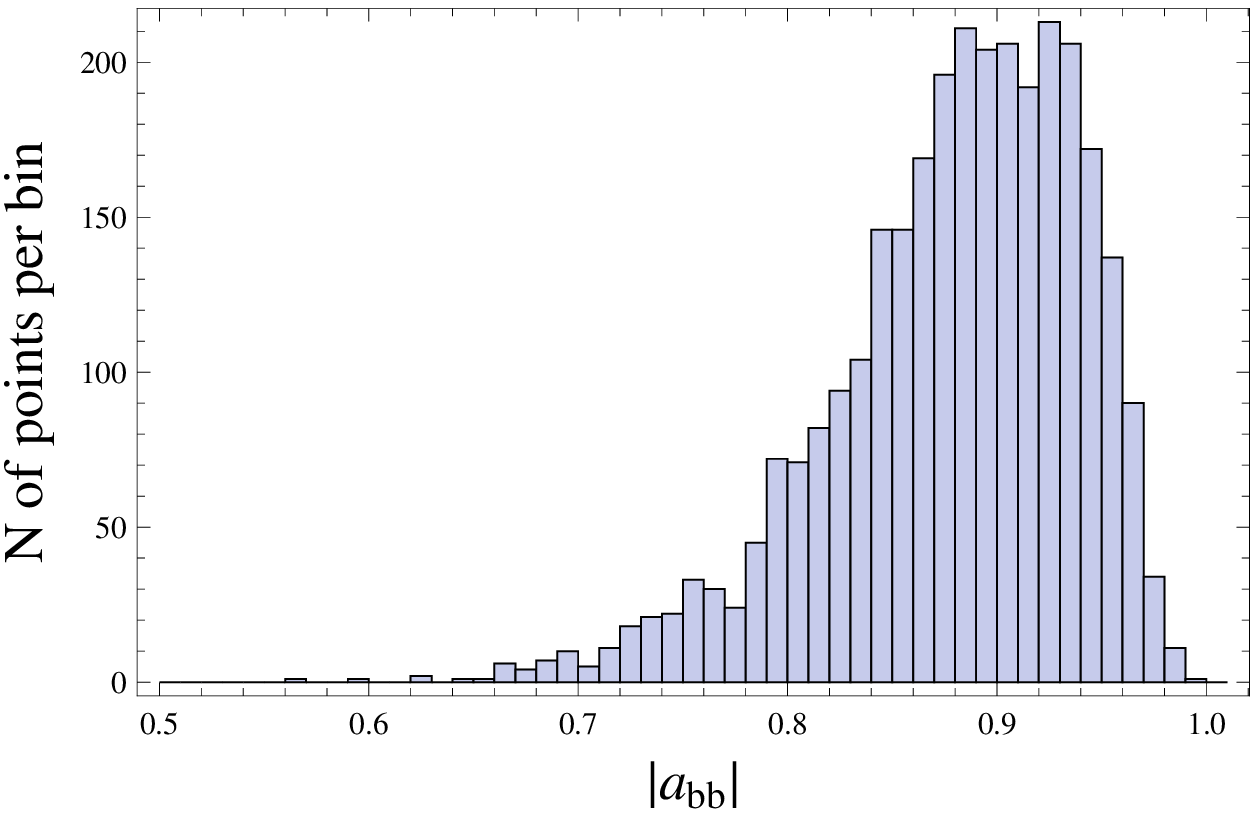}
    \includegraphics[height=5cm]{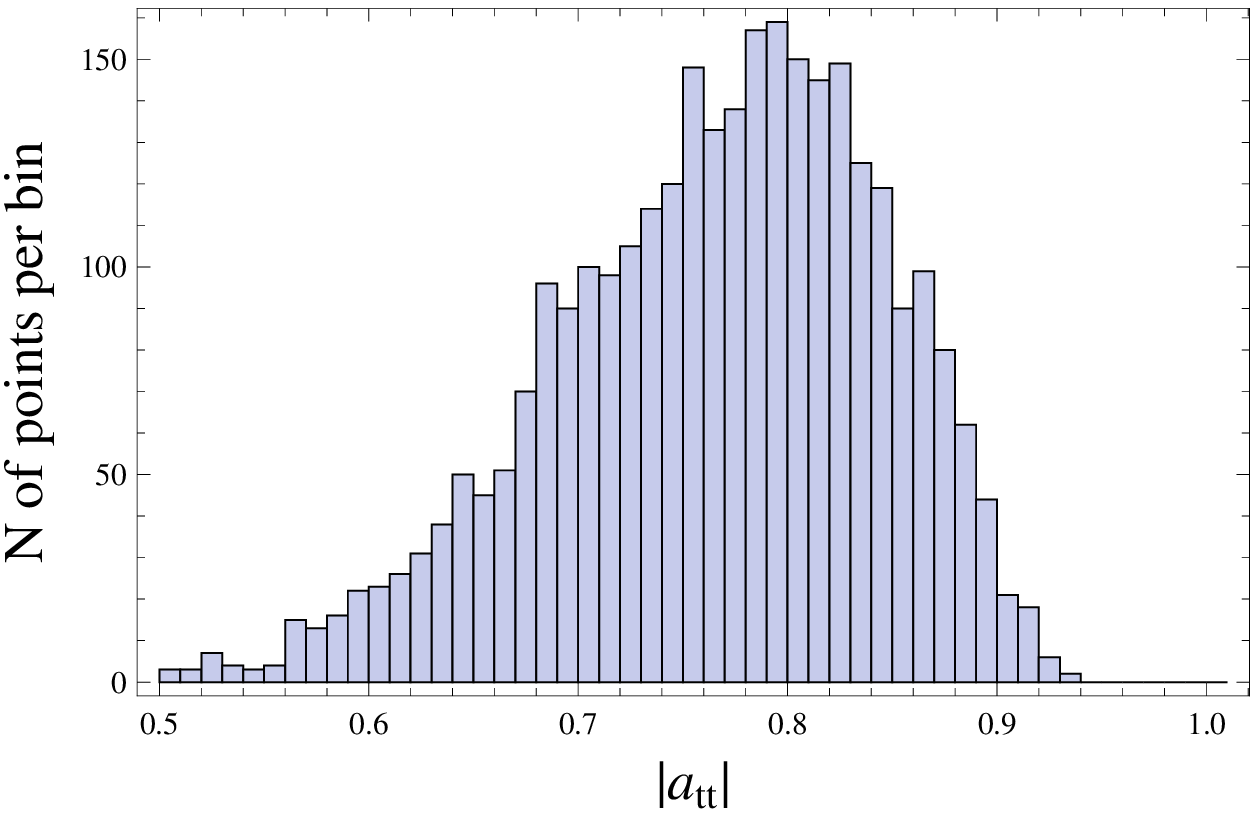}
\vspace{-.cm} \caption{Distribution of the absolute value of the
normalized Higgs couplings to $t\bar{t}$ and $b\bar{b}$, $a_{tt}$
and $a_{bb}$, in our numerical scan, with a fixed KK scale of
$R'^{-1}=1500$ GeV (KK gluon mass $M_{KKG} = 2.45 R'^{-1}$) and for
5D Yukawa couplings $|Y_{5D}^{ij}| \in [0.3,3]$. The expected
generic suppression for both couplings is demonstrated numerically
quite clearly. } \label{abbatt}
\end{figure}

We did a numerical scan over the input parameters $(Y^{5D}_1)_{ij}$,
$(Y^{5D}_2)_{ij}$, $c_{q_i}$, $c_{d_i}$, $c_{u_i}$ and we set
$R'^{-1} = 1.5$ TeV. In our scan, we pick the points that give the
correct SM quark masses and CKM matrix. Then we calculate the 4D
effective Yukawa couplings of Higgs with SM quarks. We present here
only the results for $|Y_{1,2}^{5D}|\in [0.3,3]$. First, we scan the
set of parameters with $Y^{5D}_1 = Y^{5D}_2$ which is motivated by
bulk Higgs. Here are the results for this case:
\begin{eqnarray}\label{scanresult}
a^d_{ij}= \left(\begin{array}{ccc} 0.99-1 &
0.006-0.019& 0.004-0.012\\
0.006-0.019&  0.96-0.99& 0.007-0.02 \\
0.042-0.10& 0.075-0.18& 0.85 -0.93
\end{array} \right)
\end{eqnarray}
\begin{eqnarray}\label{scanresult1}
a^u_{ij}= \left(\begin{array}{ccc} 0.99-1 &0.06-0.16
& 0.09-0.21\\
0.003-0.008& 0.94-0.98 &0.03-0.09\\
0.009-0.02& 0.05-0.14 & 0.71-0.82
\end{array} \right)
\end{eqnarray}
The first and second numbers are the $25\%$ and $75\%$ quantiles of
the distribution of $|a_{ij}|$ obtained from the scan (i.e. $50\%$
of all the values we obtained in the scan for each $|a_{ij}|$ lie
between these two quantiles). From the results we can see that the
values of $a_{ij}^{u,d}$ from the scan are consistent with the
estimates presented above (Eq \ref{adest} and \ref{auest}), and the
expected reduction of $h \bar{t} t$ coupling is confirmed. We can
also easily see this reduction of third generation Yukawa couplings
in Fig. \ref{abbatt}.

For the case  when $Y^{5D}_1$ and $Y^{5D}_2$ are completely
uncorrelated (Brane Higgs) we get the following results:
\begin{eqnarray}
a^d_{ij}= \left(\begin{array}{ccc}
0.99-1 &0.01-0.026
& 0.005-0.012\\
0.012-0.03& 0.98-1.01 &0.008-0.02\\
0.05-0.12& 0.07-0.2 & 0.96-1.03
\end{array} \right)
\end{eqnarray}

\begin{eqnarray}
a^u_{ij}= \left(\begin{array}{ccc}
0.98-1.01 &0.07-0.17
& 0.08-0.19\\
0.004-0.009& 0.97-1.02 &0.025-0.067\\
0.007-0.016& 0.04-0.11 & 0.9-0.99
\end{array} \right)
\end{eqnarray}
We can see that the off-diagonal terms of $a_{ij}^{u,d}$ are of the
same order as the previous case. However   the diagonal entries do
not have the suppression as in the $Y_1^{5D}=Y_2^{5D}$ case, see the
discussion in Subsection \ref{3rdgeneration}.

%%%%%%%%%%%%%%%%%%%%%%%%%%%%%%%%%%%%%%%%%%%%%%%%%%%%%%%%%%%%%%%%%%%%%%%%%%%%%%%%%%%%%%%%%%%%%%%%%%%%%
\section{Lepton sector}

Generically one can see that the same effects will lead to Higgs
flavor violation in the lepton sector, the only difference is that
in the lepton sector there are various ways to explain  the large
mixing angles and light masses  for the neutrinos
\cite{Perez:2008ee,AgasheOkuiSundrum,Agashe:2009tu}.
Now we want to look at
Higgs flavor violation in the charged lepton sector, then depending
on a given neutrino model, the left-handed charged lepton profiles
can be either hierarchical  and UV localized (i), or similar and UV
localized (ii).  The profiles of the right-handed charged leptons
are always hierarchical and localized near the UV brane. We treat
these two cases separately.
\begin{itemize}
\item Case (i) - left-handed and right-handed profiles are hierarchical.
Then the profiles should satisfy the following relations:
\begin{eqnarray}
f_L^i f_e^i \sim \frac{m_i^l}{\bar{Y} v_4},
\end{eqnarray}
where $f_{L,e}$ are profiles of the left-handed and right-handed
fields respectively, then the generational mixing is also
hierarchical
\begin{eqnarray}
(O_{L,e})^{i,j}\sim\frac{f_{L,e}^i}{f_{L,e}^j},\qquad~ i<j.
\end{eqnarray}
We again parameterize our Lagrangian in the following form:
\begin{equation}
{\cal{L}}_{HFV} = a^l_{ij}\sqrt{\frac{m^l_i m^l_j}{v_4^2}} H
\bar{L}^i e^j + h.c.
\end{equation}
Where $L,e$ are $SU(2)_L$ doublets and singlets respectively Then we
can estimate $a^l_{ij}$
\begin{eqnarray}
a^l_{ij}\sim \frac{2}{3}\bar{Y}^2 (v_4^2 {R'}^2)
\sqrt{\frac{f_L^if_e^j}{f_L^jf_e^i}}
\end{eqnarray}
One can see that our estimate depends on the profiles of the fermions,
but the following relation will be valid
\begin{eqnarray}
\label{hlfv1} \sqrt{|a^l_{ij}|^2+|a^l_{ji}|^2}\gtrsim \frac{4}{3}
\bar{Y}^2 (v_4^2 {R'}^2) =0.16\left( \frac{1500 \hbox{
GeV}}{1/R'}\right)^2\left( \frac{\bar Y}{3}\right)^2
\end{eqnarray}
This inequality is saturated when $\frac{f_L^i}{f_L^j}\sim
\frac{f_e^i}{f_e^j}\sim \sqrt{\frac{m^l_i}{m_j^l}}$, i.e., when the
hierarchy of charged lepton masses are explained equally by the
profiles of left-handed and right-handed fields.

\item Case (ii) - right-handed profiles are hierarchical and left-handed
profiles are similar $f_L^1 \sim f_L^2 \sim f_L^3$. Then the
profiles satisfy the following relations:
\begin{eqnarray}
f^i_L f_e^i &\sim& \frac{m^l_i}{\bar{Y} v_4} \nonumber\\
\frac{f_L^i}{f_L^j}&\sim& O(1) ,\qquad~i<j\nonumber\\
 \frac{f_e^i}{f_e^j}&\sim& \frac{m_i^l}{m_j^l},\qquad~i<j
\end{eqnarray}
then we can estimate the parameter $a^l_{ij}$ to be:
\begin{eqnarray}
\label{hlfv2} a^l_{ij}\sim \frac{2}{3}\bar{Y}^2 (v_4^2 {R'}^2)
\sqrt{\frac{f_e^j}{f_e^i}}\sim 0.08\left( \frac{1500 \hbox{
GeV}}{1/R'}\right)^2\left( \frac{\bar
Y}{3}\right)^2\sqrt{\frac{m_j^l}{m_i^l}}
\end{eqnarray}
These flavor violating Higgs Yukawa couplings to leptons can also lead to
interesting collider signals, which will also be discussed in the next section.
\end{itemize}

%%%%%%%%%%%%%%%%%%%%%%%%%%%%%%%%%%%%%%%%%%%%%%%%%%%%%%%%%%%%%%%%%%%%%%%%%%%%%%%%%%%%%%%%%%%%%%%%%%%%%
\section{Phenomenology}

The FCNC generated by flavor violating Higgs Yukawa couplings will
affect many low energy observables and also give possible signature
at colliders. In this section, we first discuss bounds  on Higgs
flavor violation coming from $\Delta F = 2$ processes such as
$\bar{K} - K$, $\bar{B} - B$, $\bar{D} - D$ mixing. And then we
discuss possible signature at the LHC including suppression of $htt$
coupling, rare top decay $t \to h c$ and flavor violating Higgs
decay $h \to \tau \mu$.

\subsection{Bounds from low energy physics}

\begin{figure}[h]
\vspace{-.2cm} \center
\includegraphics[scale=.7]{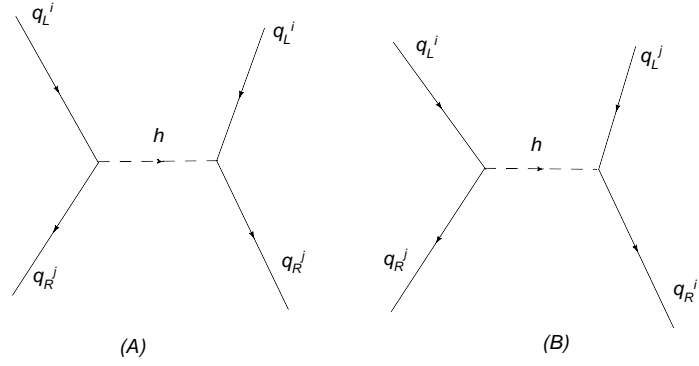}
\vspace{-.2cm} \caption{Contribution to $\Delta F = 2$ processes
from Higgs exchange}\label{higgsexchange}
\end{figure}

The $\Delta F = 2$ process can be described by the general
Hamiltonian \cite{BonadeltaF,BurasWeakHamiltonian}
\begin{eqnarray}
{\cal H}_{eff}^{\Delta F =2} = \sum_{a=1}^{5} C_a Q_a^{q_i q_j} +
\sum_{a=1}^3 \tilde{C}_a \tilde{Q}_a^{q_i q_j}
\end{eqnarray}
with
\begin{eqnarray}
Q_1^{q_i q_j} &=& \bar{q}^\alpha_{jL}\gamma_\mu
q_{iL}^\alpha\bar{q}^\beta_{jL}\gamma^\mu q^\beta_{iL},\\ \nonumber
Q_2^{q_i q_j} &=& \bar{q}^\alpha_{jR} q_{iL}^\alpha
\bar{q}^\beta_{jR} q_{iL}^\beta, \\ \nonumber Q_3^{q_i q_j} &=&
\bar{q}^\alpha_{jR}q_{iL}^\beta \bar{q}_{jR}^\beta q_{iL}^\alpha,\\
\nonumber Q_4^{q_i q_j} &=& \bar{q}^\alpha_{jR}q_{iL}^\alpha
\bar{q}_{jL}^\beta q_{iR}^\beta, \\ \nonumber Q_5^{q_i q_j} &=&
\bar{q}^\alpha_{jR} q_{iL}^\beta \bar{q}^\beta_{jL} q_{iR}^\alpha ,
\end{eqnarray}
where $\alpha, \beta$ are color indices. The operators $\tilde{Q}_a$
are obtained from $Q_a$ by exchange $L \leftrightarrow R$. For
$\bar{K}- K$ , $\bar{B}_d - B_d$, $\bar{B}_s - B_s$, $\bar{D} - D$
mixing, $q_i q_j = s d$, $b d$, $ b s$ and $ u c$ respectively.
Exchange of the Higgs can give rise to new contribution to $C_2$,
$\tilde{C}_2$ and $C_4$. This can be seen in Fig.
\ref{higgsexchange}, where Fig. \ref{higgsexchange}(A) gives $C_2$
and $\tilde{C}_2$, Fig. \ref{higgsexchange}(B) gives $C_4$. These
new contributions are
\begin{eqnarray}
C_2^h &=& a_{ij}^2 \frac{m_i m_j}{v^2}\frac{1}{m_h^2}\\
\tilde{C}_2^h &=& a_{ji}^2 \frac{m_i m_j}{v^2}\frac{1}{m_h^2}\\
C_4^h &=& a_{ij}a_{ji} \frac{m_i m_j}{v^2} \frac{1}{m_h^2}
\end{eqnarray}
where $m_h$ is the mass of physical Higgs. The model independent
bound on the new physics contribution to these Wilson coefficients
are given in \cite{BonadeltaF}.  We  use the RGE from
\cite{Bagger:1997gg} and give the bounds renormalized at the scale $\mu_h =200~ GeV$:
\begin{eqnarray}
\text{Im} C_K^2 \le \left(\frac{1}{7\times 10^7~ GeV}\right)^2,
\quad
\text{Im} C_K^4 \le \left(\frac{1}{1.3\times 10^8~ GeV}\right)^2,\\
|C_D^2| \le \left(\frac{1}{1.9\times 10^6~ GeV}\right)^2, \quad
|C_D^4| \le \left(\frac{1}{2.9\times 10^6~ GeV}\right)^2,\\
|C_{B_d}^2| \le \left(\frac{1}{0.9\times 10^6~ GeV}\right)^2,\quad
|C_{B_d}^4| \le \left(\frac{1}{1.4\times 10^6~ GeV}\right)^2,\\
|C_{B_s}^2| \le \left(\frac{1}{1\times 10^5~ GeV}\right)^2, \quad
|C_{B_s}^4| \le \left(\frac{1}{1.7\times 10^5~ GeV}\right)^2.
\end{eqnarray}
These bounds put constraints on both the Higgs flavor violating Yukawa
couplings parametrized by $a_{ij}$, and on the Higgs mass $m_h$. If we
assume that the phases of $C_{2,4}^h$ are random, i.e.,
$\text{Im}(C_{2,4}^h) \sim |C_{2,4}^h|$, we can then rewrite the
previous bounds as
\begin{eqnarray}
0.25\left(\frac{350~ GeV}{m_h}\right)^2
\frac{\text{Im}(a_{12}^d)^2}{(0.032)^2} &\le& 1, \quad
0.39\left(\frac{350~ GeV}{m_h}\right)^2
\frac{\text{Im}(a_{21}^d)^2}{(0.04)^2} \le 1, \quad
1.11\left(\frac{350~ GeV}{m_h}\right)^2 \frac{\text{Im}(a_{21}^d
a_{12}^d)}{(0.032\times0.04)} \le 1\non\\ \nonumber
0.018\left(\frac{350 ~GeV}{m_h}\right)^2
\frac{|a^u_{12}|^2}{(0.15)^2} &\le& 1,\quad 0.00005\left(\frac{350 ~
GeV}{m_h}\right)^2 \frac{|a^u_{21}|^2}{(0.008)^2} \le 1, \quad
0.0021\left(\frac{350 ~ GeV}{m_h}\right)^2
\frac{|a^u_{12}a^u_{21}|}{(0.15\times 0.008)} \le 1, \\
\nonumber
0.0002\left(\frac{350 ~ GeV}{m_h}\right)^2
\frac{|a^d_{13}|^2}{(0.01)^2} &\le& 1, \quad 0.03\left(\frac{350 ~
GeV}{m_h}\right)^2 \frac{|a^d_{31}|^2}{(0.12)^2} \le 1, \quad
0.006\left(\frac{350 ~ GeV}{m_h}\right)^2
\frac{|a^d_{13}a^d_{31}|}{(0.01\times 0.12)} \le 1\\ \nonumber
0.00003\left(\frac{350 ~ GeV}{m_h}\right)^2 \frac{|a^d_{23}|^2}{(0.01)^2}
&\le& 1, \quad 0.003 \left(\frac{350 ~ GeV}{m_h}\right)^2
\frac{|a^d_{32}|^2}{(0.15)^2} \le 1, \quad 0.001\left(\frac{350 ~
GeV}{m_h}\right)^2 \frac{|a^d_{32} a^d_{23}|}{(0.1\times 0.01)} \le
1,
\end{eqnarray}
\begin{figure}[t]
\vspace{-.2cm} \center
\includegraphics[height=7cm,width=8cm]{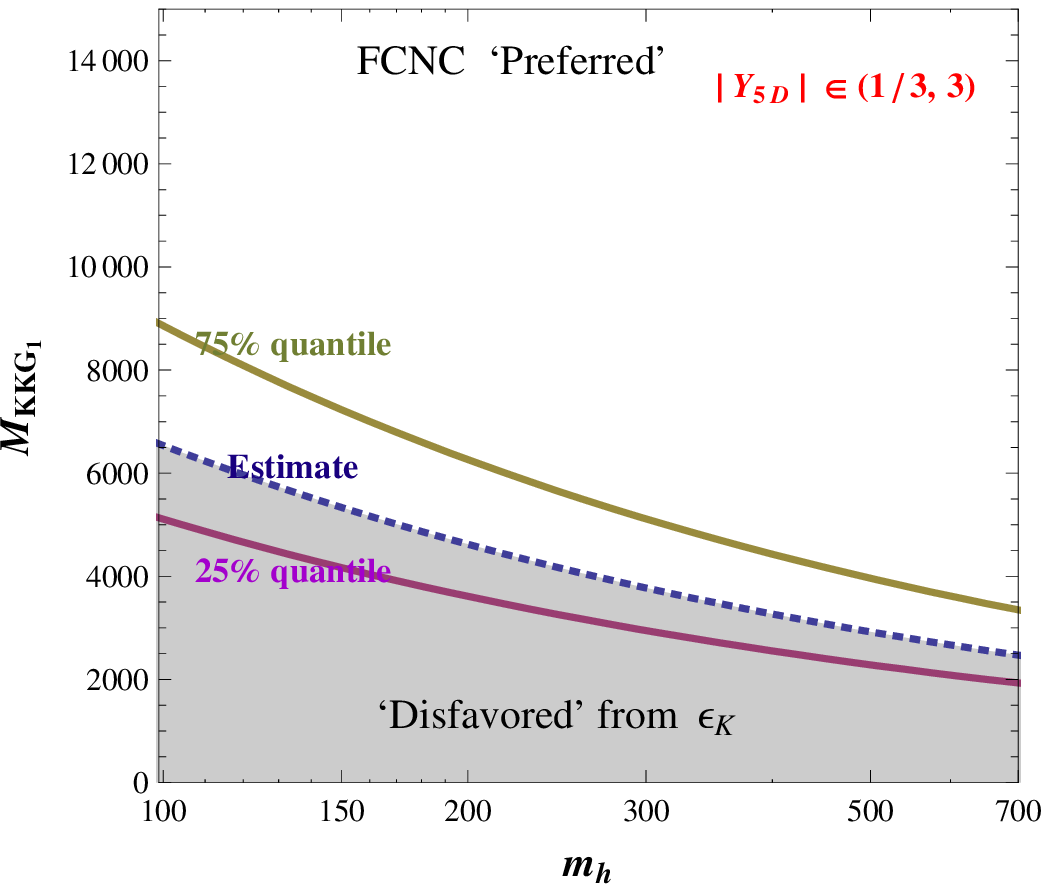}\hspace{1cm}
\includegraphics[height=7cm,width=8cm]{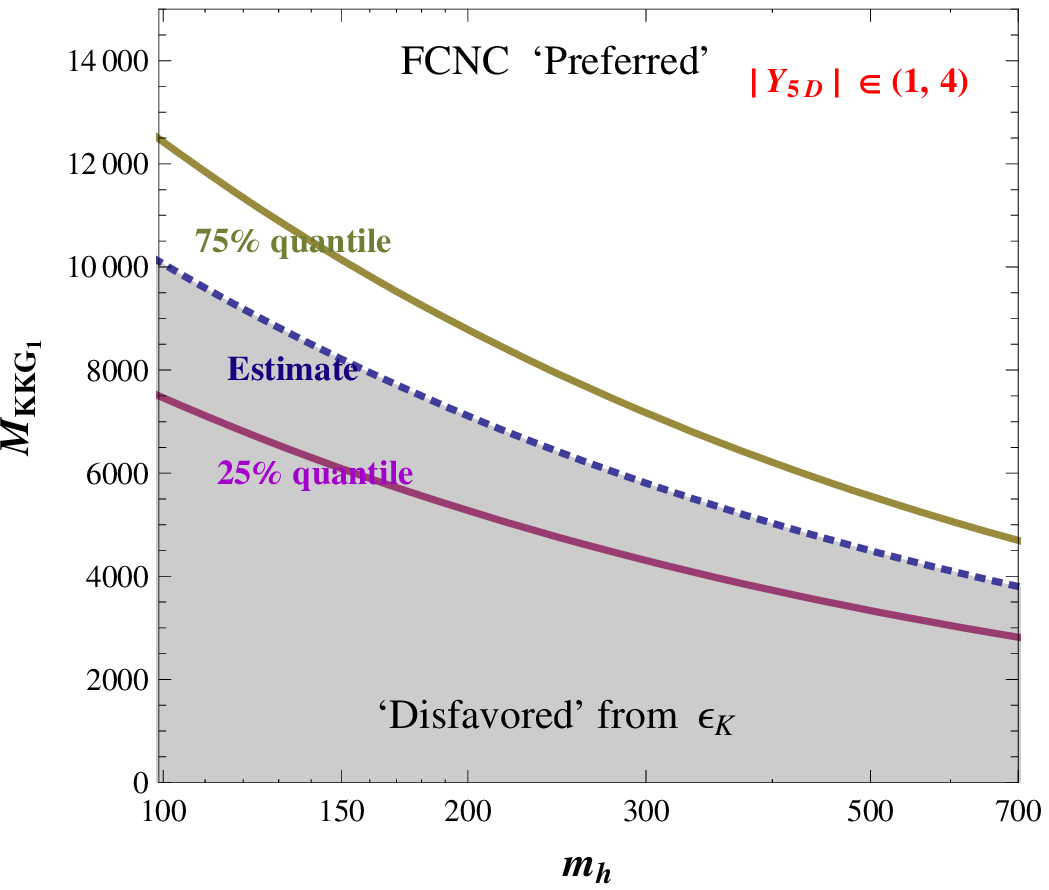}
\vspace{-.2cm} \caption{Generic bounds in the plane ($m_h,M_{KKG_1}$)
  coming from $\epsilon_K$ due to tree
level Higgs exchange, where $m_h$ is the Higgs boson mass and
$M_{KKG_1}$ is the mass of the first excited KK gluon. We perform a
scan over 5D Yukawa matrices (such that $|Y_{5D}^{ij}| \in [0.3,3]$
(left panel) and $|Y_{5D}^{ij}| \in [1,4]$ (right panel)) and over
fermion bulk c-parameters. In the scan, we choose
$Y_1^{5D}=Y_2^{5D}$ and take the $\beta \to \infty$ limit (the
result has only a mild dependence on $\beta$). The $25\%$ quantile and
$75\%$ quantile curves trace the points in this plane where $25\%$
and $75\%$ of the randomly generated parameter points are safe from
Higgs mediated FCNC's (and are otherwise in agreement with the rest
of experimental constraints in the scenario). The ``estimate'' curve
is based on the expected size of Higgs flavor violating couplings
(see Eqs.~(\ref{adest}) and (\ref{auest})) for the chosen range of
the 5D Yukawas. } \label{epsilonKbound}
\end{figure}
where we compare the $a_{ij}$ elements with their estimated values,
for a fixed average Yukawa coupling $\bar{Y}=2$ and KK scale given
by $1/R'=1500$ GeV (see formulae for the estimates from
Eqs.~(\ref{adest}) and (\ref{auest}) ). We also choose to compare
the Higgs mass with a nominal value of $m_h=350$ GeV. We can see
that the bound on $\text{Im} C_K^4$ coming from $\epsilon_K$ gives
the strongest constraint on the Higgs mass. Specifically, we have
\begin{eqnarray}
m_h  \gsim 350 ~GeV \quad \text{for}\quad \text{Im}(a_{21}^d
a_{12}^d)= (0.04\times0.032)
\end{eqnarray}
for a fixed KK scale of $1/R'= 1.5$ TeV and average 5D Yukawa of
$\bar{Y}_{5D}=2$.

In Fig.\ref{epsilonKbound}, we show the results of our numerical
scan by plotting the bounds coming from $\epsilon_K$ in the
($m_h$-$M_{KKG}$) plane, where $M_{KKG} \approx 2.45 R'^{-1}$ is the
mass of the first KK gluon. In the left panel we show results for
the case $|Y^{5D}_{ij}| \in [0.3,3]$, and
in the right panel we show results for the case $|Y^{5D}_{ij}| \in
[1,4]$. It can be seen quite clearly that a larger 5D Yukawa
coupling leads to a higher bound on the KK scale.
Note that the bounds coming from KK gluon exchange are inversely
proportional to the size of the 5D Yukawa couplings $\bar{Y}_{5D}$.
This leads to an interesting observation
\begin{itemize}
\item The new contribution to $\epsilon_K$ coming from Higgs
exchange has opposite dependence on the 5D Yukawa coupling as that
of KK gluon exchange. Thus, increasing the overall size of $Y_{5D}$
will alleviate pressure from KK gluon exchange but, as we have seen,
this will also enhance the effect of Higgs mediated FCNC's.
\end{itemize}
With the chosen $\bar{Y}_{5D}$ ($\sim 2$), we can see that for the
region of parameter space with $M_{KKG} \sim 3$ TeV (accessible at
the LHC), a Higgs mass $m_h < 400$ GeV is disfavored. On the other
hand, if a light ($< 150$ GeV) Higgs is found in the LHC, we should
expect sizable new physics contributions to $\Delta F=2$ processes,
just below current bounds.

\subsection{Collider phenomenology}

\begin{figure}[t]
\vspace{-.2cm} \center
\includegraphics[height=8cm,width=12cm]{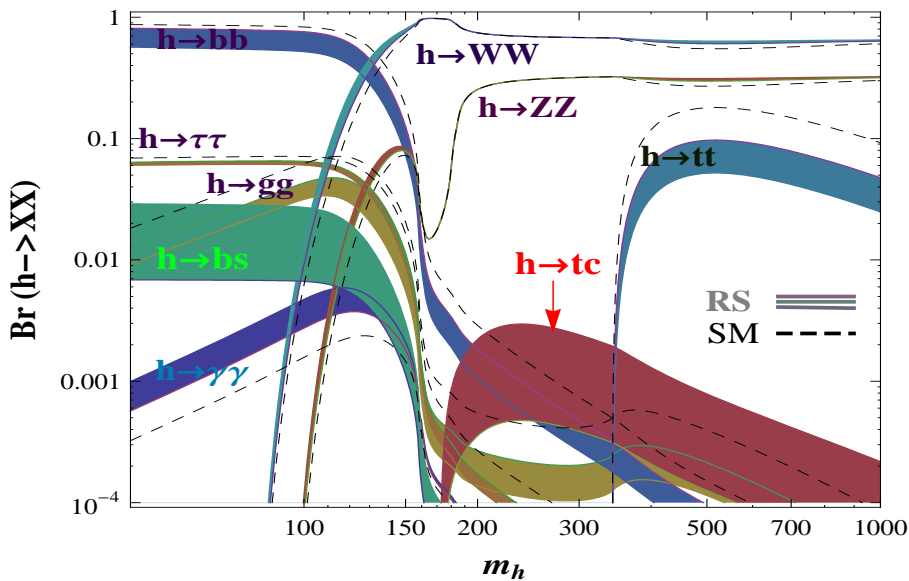}
\vspace{-.2cm} \caption{Higgs decay branching fractions as a
function of its mass, for the case of 5D Yukawas such that
$|Y_{5D}^{ij}| \in [1,4]$ and for a KK scale $R'^{-1} = 1500$ GeV
($M_{KKG_1}=2.45 R'^{-1}$). The dashed curves represent the SM
branching fractions, and the color bands correspond to $25\%$ and
$75\%$ quantiles of our scan results. The $h\to tt$ curve shows a
suppressed branching due to suppressed $htt$ couplings. This same
type of suppression happens in the $hbb$ couplings, which in turn
enhances important channels such as $h\to \gamma\gamma$. Of course
Higgs production through gluon fusion is also suppressed due to
suppressed $htt$ couplings, but vector boson fusion is assumed to
remain as in the SM, allowing one to probe at the LHC these relative
changes in the couplings. We note also the appearance of two new
important channels, $h\to bs$ and $h\to tc$, the second of which
could be looked at at the LHC if the Higgs happens to be discovered
(in the $ZZ$ channel) in the appropriate mass regime.}
\label{Branchings}
\end{figure}

Besides low energy physics constraints, there could be very
interesting signatures in colliders coming from the corrections to
the Higgs Yukawa couplings. First of all, the reduction in the $htt$
coupling, as argued in Section \ref{3rdgeneration} and confirmed by
our numerical scan, tells us that the Higgs production through gluon
fusion will be generically suppressed (at least for the bulk Higgs
scenario). This coupling can easily be suppressed by $\sim 25\%$
(for $R'^{-1}=1.5$ TeV and $\bar{Y}_{5D}\sim 2$), and therefore the
$gg\to h$ cross section will experience a reduction of  $40\%$ with
respect to the expected SM value. This reduction in Higgs events
from gluon fusion at the LHC can be observed quite clearly as well
as the relative increase in importance of the production through
gauge boson fusion \cite{CMS}. We note again that the expected
suppression is much larger in the case of a bulk Higgs, namely when
we have $Y_2=Y_1$. In the case where $Y_2$ and $Y_1$ are unrelated,
but with same overall size, there will not be a definitive
prediction on the sign of the correction to the top Yukawa and
bottom Yukawa couplings (i.e. there could be also enhancements),
although the size of the corrections is expected to be smaller than
in the $Y_2=Y_1$ case.

In the case of a light Higgs boson (and assuming that somehow low
energy FCNC bounds are overcome), the branchings of the Higgs can
change substantially due to the generically reduced $hbb$ couplings.
This would indirectly enhance the importance of $h\to\gamma\gamma$
signal, and maybe help overcome the overall reduction in the total
production cross section due to reduced top Yukawa couplings. In
Fig. \ref{Branchings}, we plot the Higgs decay branching ratio for
various final states versus the Higgs mass $m_h$ \footnote{We did
not include $h\to \mu\tau$ mode on the plot because it is model
dependent.}. We can see clearly that for a light Higgs, the
reduction in the $hbb$ coupling changes the branching ratio to other
channels significantly. For a heavy Higgs, the branching for $h\to
tt$ is reduced.

\begin{figure}[t]
\vspace{-.0cm} \center
\includegraphics[height=6.8cm,width=9.5cm]{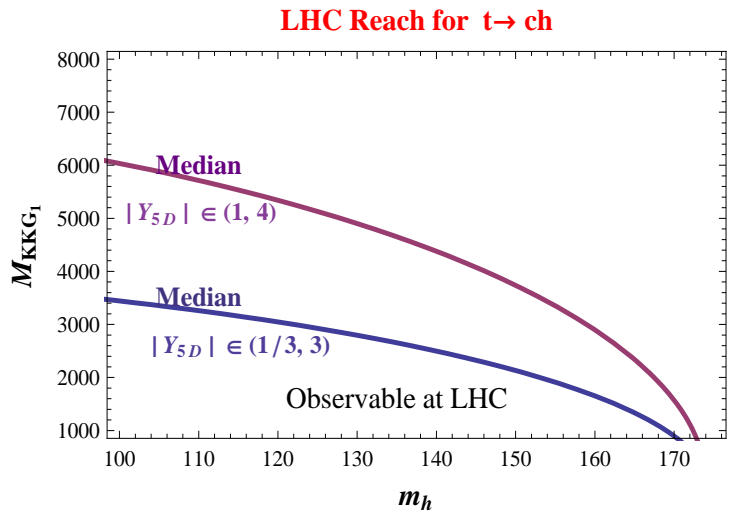}
\vspace{-.0cm} \caption{LHC observability of the exotic decay of the
  top quark $t\to ch$ in the plane ($m_h,M_{KKG_1}$). The two curves
  trace the region such that $50\%$ of the generated points in our two
  scans (one with $|Y_{5D}^{ij}| \in [0.3,3]$ and another with
  $|Y_{5D}^{ij}| \in [1,4]$) will have a visible signal at the LHC.
}
\label{tchdecay}
\end{figure}

If kinematically accessible ($m_h<m_t$), the flavor violating $htc$
couplings will allow the decay $t\to c h$ to occur. The branching
ratio of this process is given by (see for example \cite{NeubertRS})
\begin{equation}
Br(t\to c h) = \frac{2(m_t^2 - m_h^2)^2m_w^2}{(m_t^2 -
m_w^2)^2(m_t^2 + 2m_w^2)g_2^2}\left\{|a^u_{23}|^2+|a^u_{32}|^2 +
\frac{4m_c m_t}{m_t^2 - m_h^2}\text{Re}[a^u_{23}a^u_{32}]
\right\}\frac{m_c m_t}{v^2}.
\end{equation}
If we take $m_h=120$ GeV, then for $a_{23}^u \sim 0.08$ and $a_{32}^u
\sim 0.14$, which are good estimates for $\bar{Y}=2$ and a KK scale of
$1/R'=1500$ GeV (see Eq.~(\ref{auest})), we obtain a branching ratio of
\begin{equation}
Br(t\to c h) \sim 5\times 10^{-5}.
\end{equation}
The sensitivity of LHC for this rare top decay is $Br(t\to c h) \ge
6.5\times 10^{-5}$ \cite{topdecay}, precisely in the ball-park of
our estimate. In Figure \ref{tchdecay} we show the results of our
two scans, each with a different average size of the 5D Yukawas. It
is shown that observing the signal at the LHC is quite possible
although it requires larger Yukawa couplings and a  light Higgs. If
observed, this signal would be very valuable in determining the
structure of the 5D setup.

Another interesting collider signature for light Higgs might be the
Higgs lepton flavor violating decay $h\rightarrow \mu \tau$. the LHC
reach for this process was studied in \cite{Han:2000jz} and it could
be observable if $|a_{\mu\tau},(a_{\tau\mu})|>0.15$. One can see
from equations (\ref{hlfv1}) and (\ref{hlfv2}) that for case
(i), this decay is observable only for fairly large $\bar{Y}$
($\gtrsim 3$) and low KK scale $1/R' \lesssim 1.5$ TeV, while for
case (ii) there is an extra enhancement factor of
$\sqrt{\frac{m_\tau}{m_\mu}}\sim 4$ for $a_{\mu\tau}$, so that in
this case we expect larger parameter space to give us observable
effects in the $h\rightarrow \mu \tau$ decay.

For a heavy Higgs ($ m_h>m_t $), an interesting signal at the LHC
might be the Higgs flavor violating decay $h\to t c$. A similar
study on $tc$ production from radion decay was considered in
\cite{Azatov:2008vm}. From Fig. \ref{Branchings} we can see that the
branching for $h \to t c$ is in the range of $10^{-3}$ for a Higgs
mass between $200 - 300$ GeV, and for the favorable parameter values
of $\bar{Y}_{5D}\sim 2$ and $1/R'=1500$ GeV. However, even with a
branching fraction of $10^{-3}$ the signal would most likely be
dominated by large backgrounds at the LHC. Larger flavor violating
couplings are still possible for even larger values of the 5D
Yukawas, although calculability and perturbativity become then a
greater issue. More detailed analysis of the possibility and
feasibility of this channel is left for future studies.

We finally must mention that in these models one generically expects
the appearance of another light scalar in the spectrum, the radion
graviscalar. As was pointed out in \cite{Azatov:2008vm} the radion
will also typically couple to fermions with off-diagonal couplings,
and moreover, 
the two scalars could actually mix \cite{GRW} giving rise to
interesting changes in both Higgs and radion phenomenology
\cite{GRW,CGK,Korean,mtgamma}.
In that situation, the physical states emerging from the mixing will
inherit an admixture of the couplings of the
original Higgs and radion, including their off-diagonal couplings to fermions.
It would be interesting to revisit the phenomenology of Higgs-radion
mixing in view of the results obtained in this paper, although we will
leave this study for future investigations.

%%%%%%%%%%%%%%%%%%%%%%%%%%%%%%%%%%%%%%%%%%%%%%%%%%%%%%%%%%%%%%%%%%%%%%%%%%%%%%%%%%%%%%%%%%%%%%%%%%%%%
\section{Conclusion}

In this article, we computed the misalignment between Higgs Yukawa
couplings and SM fermion masses in the framework of warped extra
dimensions. We estimated this misalignment in the mass insertion
approximation and then calculated it by solving the fermion wave
functions in 5D. An important result is that the main contributions to this
misalignment are of the same order in both bulk and brane Higgs
scenarios, which means that our results are general and independent of
the Higgs localization. We first showed this fact in the bulk
Higgs case by taking the Higgs to be infinitely localized towards the IR
brane ($\beta \to \infty$); and then we also treated the brane Higgs
case, with a suitably regularized delta function localization. A subtlety in
doing the mass insertion approximation in the brane Higgs case is also
discussed.

This misalignment generally leads to FCNC mediated by the Higgs
boson. We estimated the size of these flavor changing Yukawa
couplings in models with flavor anarchy. And we confirmed our
estimates by scanning over the parameter space which reproduces the
correct quark masses and mixing angles. In addition, we found that
the Yukawa couplings of the third generation are generically
suppressed relative to their SM values.

These flavor changing Yukawa couplings  have important phenomenology
implications. First, they lead to new contributions to flavor
changing low energy observables and thus give us bounds on
parameters of the Higgs sector. We found that $\epsilon_K$ gives the
strongest bound which disfavors a light physical Higgs. In addition,
these flavor changing Yukawa couplings can give us interesting
signals at colliders. We discussed the possible reduction of Higgs
production cross section and the changes in Higgs decay branchings
at the LHC (including interesting new decay channels such as $h \to
\mu\tau$ and $h\to t c$). Another interesting signal is the rare top
decay $t\to c h$. We found that in a sizable part of the parameter
space this decay can be seen at the LHC.

\acknowledgments We would like to thank Raman Sundrum for interesting discussions,
and specially Kaustubh Agashe for his encouragement, comments and
suggestions. A.A. was partially funded by NSF No. PHY-0652363.

%%%%%%%%%%%%%%%%%%%%%%%%%%%%%%%%%%%%%%%%%%%%%%%%%%%%%%%%%%%%%%%%%%%%%%%%%%%%%%%%%%%%%%%%%

\appendix

%%%%%%%%%%%%%%%%%%%%%%%%%%%%%%%%%%%%%%%%%%%%%%%%%%%%%%%%%%%%%%%%%%%%%%%%%%%%%%%%%%%%%%%%%%%%%%%%%%%%%
\section{General misalignement formulae}\label{appendixI}

Here we present the result for the misalignment
for general fermions (both UV  and IR localized). The largest
contribution (second term of Eq. \ref{Delta}) is
\begin{eqnarray}
\Delta^d_{1}&&=2 m_d^3 R'^2 \frac{2+c_d -c_q
+\beta}{(1+2c_d)(1-2c_q)} \left [
\frac{\epsilon^{1+2c_d}}{3-c_d-c_q+\beta}
-\frac{1}{4+c_d-c_q+\beta}-{\frac{\epsilon^{2-2c_q+2c_d}}{3-c_d-c_q+\beta}}+\frac{\epsilon^{-2c_q+1}}{3+c_d+c_q+\beta}
\right.\nonumber\\
&&\left.
-\frac{\epsilon^{-2c_q+2c_d+2}}{4+2\beta}(\epsilon^{-1-2c_d}-1)
-\frac{\epsilon^{-2c_q+2c_d+2}}{4+2\beta}(\epsilon^{-1+2c_q}-1)
+\frac{\epsilon^{2c_d+1}}{5-2c_q+2\beta}(\epsilon^{-1-2c_d}-1)\right.
\nonumber\\
&&\left.
+\frac{\epsilon^{-2c_q+1}}{5+2c_d+2\beta}(\epsilon^{-1+2c_q}-1)
+\frac{\epsilon^{2+2c_d-2c_q}}{6+c_d-c_q+3\beta}(\epsilon^{-1-2c_d}-1)(\epsilon^{-1+2c_q}-1)
\right ].
\end{eqnarray}
For the case of the UV localized fermions ($c_q>0.5,c_d<-0.5$) the
3rd, 4th and 9th terms are dominating and we recover Eq. (\ref{y3}).
For the subleading contribution of the misalignment $\Delta^d_{2}$
(first term of Eq. \ref{Delta}) we get: \bea
\Delta^d_{2}=\frac{m_d^3 R'^2}{1-2c_q}\left[
-\frac{1-\epsilon^2}{\epsilon^{2c_q-1}-1}+
\frac{\epsilon^{2c_q-1}-\epsilon^2}{(\epsilon^{2c_q-1}-1)(3-2c_q)}+
\frac{\epsilon^{1-2c_q}-\epsilon^2}{(1+2c_q)(\epsilon^{2c_q-1}-1)}\right.\nonumber\\
\left.
-\frac{1}{4+c_d-c_q+\beta}+\frac{2\epsilon^{1-2c_q}}{3+c_q+c_d+\beta}
+\frac{(\epsilon^{2c_q-1}-1)\epsilon^{1-2c_q}}{5+2c_d+2\beta}
+(c_{d,q}\leftrightarrow -c_{q,d})\right] \eea For the UV localized
fermions ($c_q>0.5,c_d<-0.5$) the 3rd, 5th and 6th terms are
important and we recover Eq. (\ref{kinetic}).

%%%%%%%%%%%%%%%%%%%%%%%%%%%%%%%%%%%%%%%%%%%%%%%%%%%%%%%%%%%%%%%%%%%%%%%%%%%%%%%%%%%%%%%%%%%%%%%%%%%%%
\section{Misalignement due to $v(z)\ne h(z)$}\label{appendixII}

In this section  we discuss the possible flavor violation coming
from the the misalignment between the physical Higgs profile and the
Higgs vev profile. The profile of the KK Higgs modes are given by
\cite{bulkhiggs}
\begin{eqnarray}
h_m(z)= B z^2(Y_{1+\beta}(m R)J_\beta(mz)+J_{1+\beta}(mR) Y_\beta(m
z)).\label{higgsprofile}
\end{eqnarray}
where the mass of the KK mode is determined by the boundary
conditions. Then for the lightest mode (physical Higgs) we can
expand the Bessel functions using $(m\ll 1/z)$
\begin{eqnarray}
h(z)=A(m_H) z^{2+\beta}\left( 1-\frac{m_H^2 z^2}{4(\beta+1)}\right)
\end{eqnarray}
where the constant $A(m_H)$ is fixed by requiring the Higgs profile
normalization. One can see that in the limit ($m_H=0$), the profiles
of the physical Higgs and the profile of its vev become proportional
to each other. Then, the normalization constants of the Higgs field
and the Higgs vev, $A(m_H)$ and $V(\beta)$ (Eq. \ref{vevprof}), will
be related by
\begin{eqnarray}
A(m_H)|_{m_H=0}\equiv A(0)=\frac{V(\beta)}{v_4}
\end{eqnarray}
and so the profile of the Higgs will be given by
\begin{eqnarray}
h(z)=A(0) z^{2+\beta}\left[1+\frac{m_H^2
R'^2}{2(4+\beta)}-\frac{m_H^2
 z^2}{4(1+\beta)}+O
\left((m_H^2 R'^2)^2 \right) \right]\\ \nonumber
=\frac{v(z)}{v_4}\left[1+\frac{m_H^2 R'^2}{2(4+\beta)}-\frac{m_H^2
z^2}{4(1+\beta)}+ O \left((m_H^2 R'^2)^2 \right)\right].
\end{eqnarray}
This will lead to a new contribution to the shift $\Delta^d$
\begin{eqnarray}
\Delta^d_{3}=-m_d(m_{H}^2 R'^2) \left[
\frac{1}{2(4+\beta)}-\frac{2+\beta+c_d-c_q}{4(1+\beta)(4+\beta+c_d-c_q)}
\right],\label{deld3}
\end{eqnarray}
but one can see that in the limit $\beta\rightarrow \infty$ this
contribution decouples. Moreover, even for finite $\beta$, the
numerical size of this type of flavor misalignment is small.

%%%%%%%%%%%%%%%%%%%%%%%%%%%%%%%%%%%%%%%%%%%%%%%%%%%%%%%%%%%%%%%%%%%%%%%%%%%%%%%%%%%%%%%%%%%%%%%%%%%%%
\section{Convergent infinite sum in the mass insertion approximation}
\label{appendixIII}

In this appendix, we address again the ``contradiction'' between the mass insertion
approximation and the 5D calculation when the Higgs is on the IR brane. We
will prove that one can obtain the result of Eq. \ref{deltabrane} from
direct calculations of the Feynman diagrams in the insertion approximation.

Naively, the importance of the $Y_2$ term looks counterintuitive
because the profiles $q_R, d_L$ do vanish at IR brane. Indeed if one
follows the insertion approximation (see Fig.~\ref{insertion}) then
the coupling between $q^{KK}_R,d^{KK}_L$ and the Higgs vanish, so
there will be no contribution to fermion masses and Yukawa couplings
out of that diagram. However there is a subtlety in this approach,
since we are expanding in KK modes by using the profiles for the
case $\langle H\rangle=0$. This means that after electroweak
symmetry breaking, we should include the mixing between the whole
tower of KK modes induced by a nonzero Higgs vev. Naively the
heavier KK modes should decouple so that their contribution should
not qualitatively affect the final result. But this appears not to
be the case.

For simplicity we will start our discussion from the case of a flat
extra dimension. Now, the fermion profiles are given by sine and
cosine functions instead of Bessel functions, and the derivation becomes much
more transparent. At the same time when the Higgs is localized on
one of the branes, we still have the same issue for any Yukawa
coupling between odd modes and the Higgs i.e., the term $Y_2 q_R d_L$
naively should not lead to any misalignment between fermion masses
and Yukawa couplings.

The profiles of the even KK modes are given by
\begin{eqnarray}
q_L^n(d_R^n)&=&\frac{1}{\sqrt{\pi R}} \text{cos} \left(\frac{n z}{R
}\right), \qquad
n=\pm 1,\pm 2,...\nonumber\\
q_L^0(d_R^0)&=&\frac{1}{\sqrt{2\pi R}}
\end{eqnarray}
and the odd KK mode profiles are
\begin{eqnarray}
q_R^n&=&\frac{1}{\sqrt{\pi R}} \text{sin} \left(\frac{ n z}{R}\right)\qquad n=\pm 1,\pm2,...\nonumber\\
d_L^n&=&-\frac{1}{\sqrt{\pi R}} \text{sin} \left(\frac{ n
z}{R}\right)\qquad n=\pm 1,\pm2,...
\end{eqnarray}
The coupling $Y_2 H Q_R D_L \delta(y-\pi R)$ should vanish because
$Q_R$ and $D_L$ are vanishing at $y=\pi R$, but in the diagram
(Fig. \ref{insertion}) we have to include all the KK modes, so we will
have an infinite sum of zeroes, and in order to treat all the
infinities accurately we will again use the rectangular regulator
Eq.(\ref{regulator}) for the delta function.

Let us define the following quantities:
\begin{eqnarray}
Y^e_{mn} -\text{coupling between ``m'' and ``n'' even KK  modes} \nonumber\\
Y^o_{mn} -\text{coupling between ``m'' and ``n'' odd KK  modes}
\end{eqnarray}
then
\begin{eqnarray}
Y^e_{mn}&=&\frac{(-1)^{m+n}}{2\pi \varepsilon }\left[ \frac{\sin
\left(\frac{ (n-m) \varepsilon}{R}\right)}{n-m}+\frac{\sin\left
(\frac{(n+ m)
   \varepsilon}{R}\right)}{n+m}\right]=\frac{(-1)^{n+m}}{2\pi
R}\left[1+O\left((n,m)^2
\left(\frac{\varepsilon}{R}\right)^2\right)\right],
\nonumber\\
Y^o_{mn}&=&-\frac{(-1)^{m+n}}{2\pi \varepsilon }\left[ \frac{\sin
\left(\frac{ (n-m) \varepsilon}{R}\right)}{n-m}-\frac{\sin\left
(\frac{(n+ m)
   \varepsilon}{R}\right)}{n+m}\right]=-\frac{(-1)^{n+m}}{3\pi R}
\left(\frac{\varepsilon}{R}\right)^2 mn\left[1+ O\left((n,m)^4
\left(\frac{\varepsilon}{R}\right)^4\right)\right]
\end{eqnarray}
In a similar way one can calculate the coupling between the $0$ and the
$n$-th even KK modes:
\begin{eqnarray}
Y^e_{0n}=Y_1\frac{(-1)^{n}}{\pi\sqrt{2} \varepsilon} \frac{\sin
\left(\frac{ n
 \varepsilon}{R}\right)}{n}=Y_1\frac{(-1)^{n}}{\pi\sqrt{2} R}\left[1+O\left(
\frac{n\varepsilon}{R}\right)\right]
\end{eqnarray}
As we said before to find the $O(v^3 R'^2)$ misalignment between
fermion masses and Yukawa couplings, it is sufficient to consider
the contribution of the diagram with three Higgs insertions (see
Fig.~\ref{insertion}) and sum over all KK modes. However, for KK
modes with $|n|,|m| \gtrsim R/\varepsilon$, the sinusoidal
oscillation of the odd wavefunction inside the Higgs profile will
tend to make the $Y^o_{m,n}$ coupling vanish. Thus we need to sum up
$|n|,|m|$ only up to $\sim R/\varepsilon$, and the estimate of that
sum will be:
\begin{eqnarray}
\label{sum}
\Delta^d_{1}&\sim& v^2 \sum_{|n|,|m| =  1}^{R/\varepsilon} Y^e_{0n} \frac{R}{n}Y^o_{nm}\frac{R}{m}Y^e_{0m}\nonumber\\
&\sim& \frac{Y_1^2 Y_2 v^2}{R}\sum_{n,m  = 1}^{R/\varepsilon}\left(\frac{\varepsilon}{R}\right)^2 \nonumber\\
\end{eqnarray}
One can see that all of the terms up to $n \lesssim R/\varepsilon$
are of the same order, and so the sum should be finite and proportional
to $\frac{Y_1^2 Y_2 v^2}{R}$.
Exact resummation gives us
\begin{eqnarray}
\Delta^d_{1} = \frac{Y_1^2 Y_2 v^3}{6\pi R}
\end{eqnarray}
It is important to mention that to account for the flavor mixing
effects one has to sum at least the first $R/\varepsilon$ terms. And the
lightest mode is an admixture of the zero mode and the first
$R/\varepsilon$ KK modes. This should not be surprising because the zero
Higgs vev expansion should include all KK modes up to the value of
the cutoff and the cutoff is related to the inverse of the Higgs
wavefunction width. In our case the width of the Higgs profile is $\varepsilon$ so we have
to sum all the modes with masses up to $1/\varepsilon$.

In the case of the warped geometry things become a little bit more
complicated, because the sine and cosine are replaced by the Bessel
functions:
\begin{eqnarray}
f^e(z,m_n)&=&(Rz)^{5/2}\frac{1}{N\sqrt{R \ln(R'/R)}}\left[
J_\alpha(m_n
z)+b_\alpha(m_n)Y_\alpha(m_n z)\right]\nonumber\\
f^e(z,m_n)&=&(Rz)^{5/2}\frac{1}{N\sqrt{R \ln(R'/R)}}\left[
J_{\alpha-1}(m_nz)+b_\alpha(m_n)Y_{\alpha-1}(m_n z)\right]
\end{eqnarray}
where
\begin{eqnarray}
\alpha&=&c+\frac{1}{2}\nonumber\\
b_\alpha({m_n})&=&\frac{J_{\alpha-1}(m_n R)}{Y_{\alpha-1}(m_n
R)}=\frac{J_{\alpha-1}(m_n R')}{Y_{\alpha-1}(m_n R')}
\end{eqnarray}
but for the cases when the mass of the KK mode is $\frac{1}{R'}\ll
m\ll\frac{1}{R}$ the expressions for the profiles simplify significantly
\begin{eqnarray}
m_n R'&\sim& \pi(n+c/2+1/2)\nonumber\\
J_\alpha(m_n z)&\sim& \sqrt{\frac{2}{\pi m_n z}} \cos (m_n z-\pi/2(c+1))\nonumber\\
J_{\alpha-1}(m_n z)&\sim& \sqrt{\frac{2}{\pi m_n z}} \cos (m_n z-\pi/2 c)\nonumber\\
\end{eqnarray}
so the ratio
\begin{eqnarray}
\frac{f^o(z,m_n)}{f^e(z,m_n)}|_{z=R'-\varepsilon}\sim \frac{\sin(m_n
\varepsilon )}{\cos (m_n \varepsilon)}\sim \sin(m_n \varepsilon )
\end{eqnarray}
and so it becomes obvious that
\begin{eqnarray}
Y^o_{n l}\sim \sin(m_n \varepsilon )\sin (m_l \varepsilon).
\end{eqnarray}
One can see that $Y^o_{nl}$ has the same dependence on the KK numbers as
in the flat case, and on the masses of the KK modes $m_n\sim \pi n/R'$
for large n, so the calculation for the warp geometry will proceed
exactly as in the flat geometry case.

There is yet another way to understand this result\footnote{We thank
Raman Sundrum for suggesting it.}.
Instead of  operator $Y_2H\overline{u}_L q_R$ we can consider the
following effective operator localized at the IR brane:
\begin{eqnarray}
\frac{Y_2 (\partial_z \overline{u}_L) (\partial_zq_R) H\
\delta(z-R')}{\Lambda^2}
\label{cutoff}
\end{eqnarray}
Then the contribution to the diagram (Fig. \ref{insertion}) will be
\begin{eqnarray}
\Delta^d_{1}&\sim&\sum_{n,l\lesssim
\frac{\Lambda}{M_{kk}}} \frac{Y_1 v}{m_n} \frac{Y_2  m_n m_l}{\Lambda^2} \frac{Y_1 v}{m_l}\nonumber\\
&\sim& \frac{Y_1^2 Y_2 v^2}{\Lambda^2}\sum_{n,l\lesssim
\frac{\Lambda}{M_{kk}}}\\ \nonumber &\sim&\frac{Y_1^2 Y_2
v^2}{M_{kk}^2}
\end{eqnarray}
and we can see that the effect of every KK mode becomes equally
important and we again have to sum up all the modes up to the value
of the cutoff $\Lambda$, obtaining a cutoff independent finite
result. On the other hand it is easily seen that this operator
corresponds to giving Higgs some finite width
$\sim\frac{1}{\Lambda}$. Indeed if will
 use the boundary
conditions for the profiles $u_L|_{{}_{R'}}=q_R|_{{}_{R'}}=0$ we will get
\begin{eqnarray}
 -\left. \frac{\partial_z \overline{u}_L}{\Lambda}\right|_{R'}=
 \left(\overline{u}_L -\frac{\partial_z
   \overline{u}_L}{\Lambda}\right)_{R'}=\overline{u}_L\left(R'-\frac{1}{\Lambda}\right)+O\left(\frac{1}{\Lambda^2}
 \right)
\end{eqnarray}
so the operator (\ref{cutoff}) is equivalent to
\begin{eqnarray}
\frac{(\partial_z \overline{u}_L) (\partial_zq_R) H
\delta(z-R')}{\Lambda^2}\Leftrightarrow (\overline{u}_L  q_R) H\delta\left(z-R'-\frac{1}{\Lambda} \right)
\end{eqnarray}
This result is not surprising because the width of the Higgs profile
should be related to the value of the inverse cutoff.

\end{document}